\documentclass[10pt, twocolumn]{IEEEtran}
\usepackage{color}
\usepackage{float}
\usepackage{cite}
\usepackage{amsmath}
\usepackage{amsfonts}
\usepackage{amssymb}
\usepackage{algorithm}
\usepackage{algorithmic}
\usepackage{graphicx}
\usepackage{verbatim}

\newtheorem{theorem}{Theorem}
\newtheorem{lemma}{Lemma}
\newtheorem{remark}{Remark}

\newtheorem{corollary}{Corollary}
\newtheorem{example}{Example}

\newcommand{\beq}{\begin{equation}}
\newcommand{\eeq}{\end{equation}}
\newcommand{\beqnn}{\begin{equation*}}
\newcommand{\eeqnn}{\end{equation*}}
\newcommand{\beqy}{\begin{eqnarray}}
\newcommand{\eeqy}{\end{eqnarray}}
\newcommand{\beqynn}{\begin{eqnarray*}}
\newcommand{\eeqynn}{\end{eqnarray*}}
\newcommand{\bit}{\begin{itemize}}
\newcommand{\eit}{\end{itemize}}
\newcommand{\ben}{\begin{enumerate}}
\newcommand{\een}{\end{enumerate}}
\newcommand{\bex}{\begin{example}}
\newcommand{\eex}{\end{example}}

\newcommand{\balg}[1]{\begin{algorithm} \caption{#1}}
\newcommand{\ealg}{\end{algorithm}}

\newcommand{\balgc}{\begin{algorithmic}[1]}
\newcommand{\ealgc}{\end{algorithmic}}

\newcommand{\bary}{\begin{array}}
\newcommand{\eary}{\end{array}}
\newcommand{\bmx}{\begin{bmatrix}}
\newcommand{\emx}{\end{bmatrix}}
\newcommand{\bsmx}{\left[\begin{smallmatrix}}
\newcommand{\esmx}{\end{smallmatrix}\right]}
\newcommand{\bmxc}[1]{\left[\begin{array}{@{}#1@{}}}
\newcommand{\emxc}{\end{array}\right]}
\newcommand{\bcn}{\begin{center}}
\newcommand{\ecn}{\end{center}}

\newcommand{\Rbb}{{\mathbb{R}}}
\newcommand{\Zbb}{{\mathbb{Z}}}

\newcommand{\mbn}{{m \times n}}

\newcommand{\Rnbn}{\Rbb^{n \times n}}

\newcommand{\Rmbm}{\Rbb^{m \times m}}

\newcommand{\Zn}{\Zbb^{n}}

\newcommand{\sss}{\scriptscriptstyle }

\newcommand{\sOB}{{\scriptscriptstyle \text{OB}}}

\newcommand{\sL}{{\scriptscriptstyle L}}
\newcommand{\sS}{{\scriptscriptstyle S}}
\newcommand{\sV}{{\scriptscriptstyle V}}

\newcommand{\sBB}{{\scriptscriptstyle \text{BB}}}

\newcommand{\A}{\boldsymbol{A}}

\newcommand{\D}{\boldsymbol{D}}

\newcommand{\G}{\boldsymbol{G}}

\newcommand{\I}{\boldsymbol{I}}

\renewcommand{\P}{\boldsymbol{P}}
\newcommand{\Q}{\boldsymbol{Q}}
\newcommand{\R}{\boldsymbol{R}}

\newcommand{\U}{\boldsymbol{U}}
\newcommand{\V}{\boldsymbol{V}}

\newcommand{\Z}{\boldsymbol{Z}}

\newcommand{\e}{\boldsymbol{e}}

\renewcommand{\l}{\boldsymbol{\ell}}
\newcommand{\n}{\boldsymbol{n}}

\newcommand{\s}{\boldsymbol{s}}
\renewcommand{\u}{\boldsymbol{u}}
\renewcommand{\v}{\boldsymbol{v}}
\newcommand{\w}{\boldsymbol{w}}
\newcommand{\x}{{\boldsymbol{x}}}
\newcommand{\y}{{\boldsymbol{y}}}
\newcommand{\z}{\boldsymbol{z}}
\newcommand{\0}{{\boldsymbol{0}}}

\newcommand{\br}{{\bar{r}}}
\newcommand{\bby}{{\bar{\y}}}
\newcommand{\bbx}{{\bar{\x}}}

\newcommand{\bbQ}{{\bar{\Q}}}
\newcommand{\bbR}{{\bar{\R}}}
\newcommand{\bbu}{{\bar{\u}}}
\newcommand{\bbl}{{\bar{\l}}}

\newcommand{\tv}{{\tilde{v}}}
\newcommand{\ty}{{\tilde{y}}}

\newcommand{\tby}{{\tilde{\y}}}

\newcommand{\hx}{{\hat{x}}}
\newcommand{\hz}{{\hat{z}}}
\newcommand{\hbx}{{\hat{\x}}}

\newcommand{\hbz}{{\hat{\z}}}

\usepackage{slashbox}

\begin{document}

\title{Success probability of the Babai estimators for \\  box-constrained integer linear models}

\author{Jinming~Wen
        and~Xiao-Wen~Chang% <-this % stops a space
\thanks{This work was supported by NSERC of Canada grant 217191-12 and ANR through the HPAC project under Grant ANR 11 BS02 013.}
\thanks{J.~Wen was with  The Department of Mathematics and Statistics, McGill University, Montreal, QC H3A 0B9, Canada, and CNRS, Laboratoire de l'Informatique du Parall\'elisme (U. Lyon, CNRS, ENSL, INRIA, UCBL),
Lyon 69007, France. He is with  the Department of Electrical and Computer Engineering, University of Alberta, Edmonton T6G 2V4, Canada (e-mail: jinming1@ualberta.ca).}
\thanks{X.-W. Chang is with The School of Computer Science, McGill University,
Montreal, QC H3A 2A7, Canada (e-mail: chang@cs.mcgill.ca). }

\thanks{Manuscript received; revised .}}

\markboth{To appear in IEEE Transactions on Information Theory }%
{Shell \MakeLowercase{\textit{et al.}}: Bare Demo of IEEEtran.cls for Journals}

\maketitle

\begin{abstract}

In many applications including communications, one may encounter a linear model where the parameter vector $\hbx$ is an integer vector in a box.
To estimate $\hbx$, a typical method is to solve  a box-constrained integer least squares  (BILS) problem.
However, due to its high complexity, the box-constrained Babai integer point $\x^\sBB$  is commonly used
as a suboptimal solution.
In this paper, we first derive formulas for the success probability $P^\sBB$ of $\x^\sBB$ and the success probability $P^\sOB$
of the ordinary Babai integer point $\x^\sOB$ when $\hbx$ is uniformly distributed over the constraint box.
Some properties of $P^\sBB$ and $P^\sOB$ and the relationship between them are studied.
Then, we investigate the effects of some column permutation strategies on $\P^\sBB$.
In addition to V-BLAST and SQRD, we also consider the permutation strategy involved in the  LLL lattice reduction, to be referred to as LLL-P.
On the one hand, we show that when the noise is relatively small, LLL-P always  increases $P^\sBB$ and argue why both V-BLAST and SQRD often increase $P^\sBB$;
and on the other hand, we show that when the noise is relatively large, LLL-P always decreases $P^\sBB$ and argue why both V-BLAST and SQRD often decrease $P^\sBB$.
We also derive a column permutation invariant bound on  $P^\sBB$, which is an upper bound and a lower bound
under these two opposite conditions, respectively. Numerical results demonstrate our findings.
Finally, we consider a conjecture concerning   $\x^\sOB$ proposed by Ma  et al.
We first construct an example to show that the conjecture does not hold in general, and then show that it does hold
under some conditions.

\end{abstract}

\begin{IEEEkeywords}
Box-constrained integer least squares estimation, Babai integer point, success probability, column permutations, LLL-P,
SQRD, V-BLAST.
\end{IEEEkeywords}

\section{Introduction}
\IEEEPARstart{S}{uppose} that we have the following box-constrained linear model:
\begin{subequations}
\label{e:lm0}
\begin{align}
& \y=\A\hbx+\v, \quad \v \sim \mathcal{N}(\boldsymbol{0},\sigma^2 \I) \label{e:model} \\
& \hbx\in \mathcal{B} \equiv \{\x \in \mathbb{Z}^{n}:\l \leq \x\leq \u,\; \l, \u\in \mathbb{Z}^{n}\}
\label{e:box}
\end{align}
\end{subequations}
where $\y\in \mathbb{R}^m$ is an observation vector,   $\A\in\mathbb{R}^{m\times n}$ is a deterministic
model matrix with full column rank, $\hbx$ is an unknown integer parameter vector in the box $\mathcal{B}$,
$\v\in \mathbb{R}^m$ is a noise vector following the Gaussian distribution
$\mathcal{N}(\boldsymbol{0},\sigma^2 \I)$ with $\sigma$ being known.
This model arises in various applications including wireless communications, see e.g., \cite{DamGC03}, \cite{FisW03}.
In this paper, we assume that  $\hbx$ is random and  uniformly distributed over the box $\mathcal{B}$.
This assumption is often made for MIMO applications, see, e.g., \cite{JalO05}.

A common method to estimate/detect  $\hbx$ in \eqref{e:lm0} is to solve the following box-constrained integer least squares (BILS) problem:
\beq
\label{e:BILS}
\min_{\x\in\mathcal{B}}\|\y-\A\x\|_2^2
\eeq
whose solution is the maximum likelihood estimator/detector of $\hbx$.
Here we would like to make a comment on terminology.
In communications, it is proper to use ``detect'' and ``detector'' for the constrained case.
However, later in this paper we will use
``estimate'' and ``estimator'' as an extension of the terminology commonly used in the unconstrained case.
A typical approach to solving \eqref{e:BILS} is discrete search, which usually consists of two stages: reduction and search.
In the first stage, orthogonal transformations are
used to transform $\A$ to an upper triangular matrix $\R$.
To make the search process  more efficient, a column permutation strategy  is often  used in reduction.
Two well-known strategies are V-BLAST \cite{FosGVW99}, \cite{DamGC03} and SQRD \cite{WubBRKK01}, \cite{ChaH08}.
The commonly used search methods are the so-called sphere decoding methods \cite{DamGC03}, \cite{BouGBF03} and \cite{ChaH08},
which are the extensions of the Schnorr-Euchner search method  \cite{SchE94}, a variation of the Fincke-Pohst search method \cite{FinP85},
for ordinary integer least squares problems to be mentioned below.
There are also some variants of Schnorr-Euchner search methods, see, e.g., \cite{WenZMC16}.

If the true parameter vector $\hbx \in \Zn$ in the linear model \eqref{e:model} is not subject to any constraint, then
we say \eqref{e:model} is an ordinary linear model.
In this case, to estimate $\hbx$, one
solves an ordinary integer least squares (OILS) problem (also referred to as the closest vector problem):
\beq
\label{e:OILS}
\min_{\x\in\mathbb{Z}^n}\|\y-\A\x\|_2^2
\eeq
and whose solution is referred to as the OILS estimator of $\hbx$.
Algorithms and theory for OILS problems are surveyed in \cite{AgrEVZ02} and  \cite{HanPS11}.

The most widely used reduction strategy  in solving \eqref{e:OILS} is  the LLL reduction \cite{LenLL82},
which consists of two types of operations called size reduction and column permutation.
But it is difficult to use it to solve a BILS problem because after size reductions
the box constraint becomes too complicated  to handle in the search process.
However, one can use  its permutation strategy, to be referred to as LLL-P (we referred it to as LLL-permute in \cite{ChaWX13}).
The LLL-P, SQRD and V-BLAST strategies use only the information of $\A$ to do the column permutations.
Some column permutation strategies which use not only the information of $\A$, but also the information of $\y$ and the box constraint
have also been proposed \cite{SuW05},  \cite{ChaH08} and \cite{BreC11}.

For a fixed constraint box ${\cal B}$ in \eqref{e:box}, where all the entries of $\l$ are equal and all the entries of $\u$ are equal,
it was shown in \cite{JalO05} that  when the signal-to-noise ratio (SNR) is fixed
the expected complexity of solving \eqref{e:BILS} by the Fincke-Pohst search method
behaves as an exponential function of the dimension $n$ when $n$ is large enough,
although it is dominated by polynomial terms for high SNR and small $n$ \cite{HasV05}\cite{JalO05}.
So for some real-time applications, an approximate solution, which can be produced quickly, is  computed instead.
For the OILS problem, the Babai integer point $\x^\sOB$,  to be referred to as the ordinary Babai estimator,
which can be obtained by the Babai nearest plane algorithm \cite{Bab86}, is an often used approximate solution.
Taking the box constraint into account, one can easily modify the Babai nearest plane algorithm
to get an approximate solution $\x^\sBB$ to the BILS problem \eqref{e:BILS},
to be referred to as the box-constrained Babai estimator.
This estimator is  the first point found by the search methods proposed in \cite{BouGBF03}, \cite{DamGC03} and \cite{ChaH08},
and it has been used as a suboptimal solution, see, e.g., \cite{WinFH04}.
In communications,  algorithms for finding the Babai estimators are often referred to as successive interference cancellation detectors.
There have been  algorithms which find  other suboptimal solutions to the BILS problems in communications, see, e.g.,
\cite{MaDWLC02, ArtSH03, GuoN06, GowH07, BarT08, ShiK08, JalBOT09, WuT09, LinMG09, LaiHJ11} etc.
In this paper we will focus only on the Babai estimators.

In order to see how good an estimator is, one needs to find the probability of the estimator being equal to the true integer parameter vector,
which is referred to as success probability \cite{HasB98}.
The probability of wrong  estimation is referred to as error probability, see, e.g., \cite{JalBOT09}.

For the estimation of $\hbx$ in the ordinary linear model \eqref{e:model},
where $\hbx$ is supposed to be deterministic,
the formula  of the success probability $P^\sOB$ of the ordinary Babai estimator  $\x^\sOB$
was first  given in \cite{Teu98a}, which considers a variant form of the ILS problem \eqref{e:OILS}.
A simple derivation for an equivalent formula of $P^\sOB$  was given in \cite{ChaWX13}.
It was shown in \cite{ChaWX13} that $P^\sOB$ increases after applying the LLL reduction algorithm  or only the
LLL-P column permutation strategy, but $P^\sOB$ may strictly decrease after applying the SQRD and V-BLAST permutation strategies.

The main goal of this paper is to extend the main results we obtained in \cite{ChaWX13}
for the ordinary case  to the box-constrained case.
We will present a formula  for the success probability $P^\sBB$ of the box-constrained Babai estimator $\x^\sBB$
and a formula  for the success probability $P^\sOB$ of the ordinary Babai estimator $\x^\sOB$
when  $\hbx$ in \eqref{e:lm0} follows a uniform distribution over the box $\mathcal{B}$.
Some properties of $P^\sBB$ and $P^\sOB$ and the relationship between them
will also be given.

Then  we will investigate the effect of the LLL-P column permutation strategy on $P^\sBB$.
We will show that $P^\sBB$ increases under a condition.
Surprisingly,  we will also show that $P^\sBB$ decreases after  LLL-P is applied under an opposite condition.
Roughly speaking, these two opposite conditions  are that the noise standard deviation $\sigma$ in \eqref{e:model} are relatively small
and large, respectively.
This is different from the ordinary case, where $P^\sOB$ always increases after the LLL-P strategy is applied.
Although our theoretical results for LLL-P cannot be extended to SQRD and V-BLAST,
our numerical tests indicate that under the two conditions, often (not always) $P^\sBB$ increases and decreases, respectively,
after applying SQRD or V-BLAST.
Explanations will be given for these phenomena.
These suggest that before we applying LLL-P, SQRD or V-BLAST we should check the conditions.
Moreover, we will give a bound on $P^\sBB$, which is column permutation invariant.
It is interesting that the bound is  an upper bound under the small noise condition we just mentioned
and becomes a lower  bound under the  opposite condition.

In  \cite{MaHFL09},  the authors made a conjecture, based on which  a stopping criterion for the search process
was proposed to reduce the computational cost of solving the BILS problem.
The conjecture is related to the success probability $P^\sOB$ of the ordinary Babai estimator $\x^\sOB$.
We will first  show that the conjecture does not always hold and then show it holds under a condition.

The rest of the paper is organized as follows.
In Section \ref{s:crs}, we introduce the QR reduction and the LLL-P, SQRD and V-BLAST column recording strategies.
In Section \ref{s:pba}, we present the formulas for $P^\sBB$ and $P^\sOB$,
study the properties of $P^\sBB$ and $P^\sOB$ and the relationship between them.
In Section \ref{s:proba}, we investigate the effects of the LLL-P, SQRD and V-BLAST column permutation strategies  and
derive a bound on $P^\sBB$.
In Section \ref{s:conjecture}, we investigate the conjecture made in  \cite{MaHFL09} and obtain some negative and positive results.
Finally,  we summarize this paper in Section \ref{s:sum}.

{\bf Notation}. For matrices, we use bold upper-case letters and for vectors we use bold lower-case letters.
For $\x\in \mathbb{R}^n$, we use $\lfloor \x\rceil$ to denote its nearest integer vector, i.e.,
each entry of $\x$ is rounded to its nearest integer (if there is a tie, the one with smaller magnitude is chosen).
For a vector $\x$, $\x_{i:j}$ denotes the subvector of $\x$ formed by entries $i, i+1, \ldots,j$.
For a matrix $\A$, $\A_{i:j,i:j}$ denotes the submatrix of $\A$ formed by rows and columns $i, i+1, \ldots,j$.

\section{QR Factorization and Column Reordering}\label{s:crs}

Assume that the model matrix $\A$ in the linear model \eqref{e:model} has the QR factorization
\beq
\A=[\Q_1, \Q_2]\bmx\R \\ \0 \emx
\label{e:qr}
\eeq
where $[\underset{n}{\Q_1}, \underset{m-n}{\Q_2}]\in \Rmbm$ is orthogonal and $\R\in \Rnbn$ is upper triangular.
Without loss of generality, we assume that the diagonal entries of $\R$ are positive throughout the paper.
Define $\tby=\Q_1^T\y$ and $\tilde{\v} = \Q_1^T \v$.
Then,  the linear model \eqref{e:lm0}  is reduced to
\begin{subequations}\label{e:lmqr}
\begin{align}
& \tby=\R\hbx+\tilde{\v}, \quad \tilde{\v} \sim \mathcal{N}(\0, \sigma^2 \I), \label{e:modelqr} \\
& \hbx\in \mathcal{B} \equiv \{\x \in \mathbb{Z}^{n}:\l \leq \x\leq \u,\; \l, \u\in \mathbb{Z}^{n}\}
\label{e:boxqr}
\end{align}
\end{subequations}
 and the BILS problem \eqref{e:BILS} is reduced to
\beq
\label{e:BILSR}
\min_{\x\in\mathcal{B}}\|\tby-\R\x\|_2^2.
\eeq

To solve the reduced problem \eqref{e:BILSR}, sphere decoding search algorithms are usually used to find the optimal solution.
For search efficiency, one typically adopts a column permutation strategy, such as V-BLAST, SQRD or LLL-P, in the reduction process to obtain a better $\R$.
For simplicity, we assume that the column permutations are performed on $\R$ in  \eqref{e:qr}
no matter which strategy is used, i.e.,
\beq
\label{e:permu}
\bbQ^T\R \P = \bbR
\eeq
where $\bbQ  \in \mathbb{R}^{n\times n}$ is orthogonal, $\P\in \mathbb{Z}^{n\times n}$ is a permutation matrix,
and  $\bbR\in \mathbb{R}^{n\times n}$ is an upper triangular matrix
satisfying the properties of the corresponding column permutation strategies.
Notice that combining \eqref{e:qr} and \eqref{e:permu} result in the following QR factorization of
the column reordered $\A$:
$$
\A \P = \tilde{\Q}\bmx \bbR \\ \0 \emx, \  \ \tilde{\Q}\equiv \Q \bmx \bbQ & \0 \\  \0 & \I_{m-n} \emx .
$$

The V-BLAST strategy determines the columns of  $\bbR$  from the last to the first.
Suppose columns $n, n-1, \ldots, k+1$ of $\bbR$ have been determined, this strategy chooses a column from $k$ remaining
columns of $\R$ as the $k$-th column such that $\br_{kk}$  is maximum over all of the $k$ choices.
For more details, including efficient algorithms, see \cite{FosGVW99, DamGC03, Has04, ChaP07, ZhuCLG11} etc.
One may refer to \cite{LoyG04} for the performance analysis of V-BLAST.

In contrast to V-BLAST, the SQRD strategy determines the columns of $\bbR$ from the first to the last
by using the modified Gram-Schmidt algorithm or the Householder QR algorithm.
Suppose columns $1,2,\ldots,k-1$ of $\bbR$ have been determined.
In the $k$-th step of the algorithm, the $k$-th column of  $\bbR$ we seek is chosen from the remaining $n-k+1$ columns of $\R$
such that $\br_{kk}$  is smallest.
For more details, see \cite{WubBRKK01} and \cite{ChaH08} etc.

The LLL-P strategy \cite{ChaWX13}  does the column permutations of the LLL reduction algorithm
and produces  $\bbR$ satisfying  the Lov\'{a}sz condition:
\beq
\label{e:lovasz}
\delta\,\br_{k-1,k-1}^2 \leq  \br_{k-1,k}^2+ \br_{kk}^2,\quad k=2, 3, \ldots, n
\eeq
where $\delta$ is a parameter satisfying $1/4 < \delta \leq 1$.
Suppose that  $\delta\, r_{k-1,k-1}^2 > r^2_{k-1,k}+r^2_{k,k}$ for some $k$.
Then we interchange columns $k-1$ and $k$ of $\R$.
After the permutation,  the upper triangular structure of $\R$ is no longer maintained.
But we can bring $\R$ back to an upper triangular matrix by using the Gram-Schmidt orthogonalization technique
(see \cite{LenLL82}) or by a Givens rotation:
\beq
\label{e:lllpermu}
\bbR=\G_{k-1,k}^T\R\P_{k-1,k}
\eeq
where $\G_{k-1,k}$ is an orthogonal matrix and $\P_{k-1,k}$ is a permutation matrix, and $\bbR$ satisfies
\beq
\begin{split}
\br_{k-1,k-1}^2&= r^2_{k-1,k}+r^2_{k,k}, \\\ \
\br^2_{k-1,k}+\br^2_{k,k}&= r_{k-1,k-1}^2,\\ \ \
\br_{k-1,k-1}\br_{kk}&=r_{k-1,k-1}r_{kk}.
\label{e:rbr}
\end{split}
\eeq
Note that the above operation  guarantees that the inequality in \eqref{e:lovasz} holds.
For simplicity, later when we refer to a column permutation, we mean the whole process
of a column permutation and triangularization.
For readers' convenience, we describe the  LLL-P strategy  in Algorithm \ref{a:LLL},
which can also be called the LLL-P reduction.

\begin{algorithm}[h!]
\caption{LLL-P}   \label{a:LLL}
\begin{algorithmic}[1]
%  \STATE compute the QR factorization: $\A=\Q\bmx \R \\ \0 \emx$;
  \STATE set $\P=\I_n$, $k=2$;
  \WHILE{$k\le n$}
   \IF{$ \delta\, r_{k-1,k-1}^2>  r^2_{k-1,k}+r^2_{kk}$}
    \STATE perform a column permutation:
    \; $\R\!=\!\G^T_{k-1,k}\R\P_{k-1,k}$; \label{l:pt}
    \STATE update $\P$: $\P=\P\P_{k-1,k}$;
    \STATE $k=k-1$, when $k>2$;
   \ELSE
    \STATE $k=k+1$;
   \ENDIF
  \ENDWHILE
%  \STATE set $\bbR=\R$;
\end{algorithmic}
\end{algorithm}

Here we give a remark about the LLL-P algorithm.
Note that  the LLL-P algorithm is the same as the original LLL algorithm,  except that any operations related to
size reductions are not performed.
When the Lov\'{a}sz condition  \eqref{e:lovasz}  for  two consecutive columns $k-1$ and $k$ of $\R$ is not satisfied,
the algorithm interchanges the two columns and performs triangularization.
We have just shown that the two updated columns satisfy the Lov\'{a}sz condition.
The algorithm terminates when the Lov\'{a}sz condition for any two consecutive columns is satisfied.
The  proof for the convergence of the original LLL algorithm,
which does not use the size reduction condition,  can be applied here to show
the convergence of the LLL-P algorithm.
We would like to point out that as the size reduction condition ($|r_{ij}| \leq r_{ii}/2$) in the LLL reduction
is not satisfied any more, some properties of the LLL reduction are lost in the LLL-P reduction.

With the QR factorization \eqref{e:permu}, we define
\beq
\label{e:permtrans}
\begin{split}
& \bby=\bbQ^T\tby, \ \ \hbz=\P^T \hbx, \ \  \bar{\v}= \bbQ^T\tilde{\v}, \\
& \z=\P^{T}\x, \ \ \bar{\l}=\P^T\l, \ \ \bar{\u}=\P^T\u.
\end{split}
\eeq
Then the linear model \eqref{e:lmqr} is transformed to
\begin{subequations} \label{e:nlmqr}
\begin{align}
& \bby = \bbR \hbz + \bar{\v},  \ \ \bar{\v} \sim \mathcal{N}(\0, \sigma^2 \I), \label{e:nmodelqr} \\
& \hbz \in \bar{\mathcal{B}}=\{\z\in \mathbb{Z}^{n}:\bar{\l}\leq \z\leq \bar{\u}, \ \bar{\l}, \ \bar{\u}\in \mathbb{Z}^n\}
 \label{e:nboxqr}
\end{align}
\end{subequations}
and  the BILS problem \eqref{e:BILSR} is transformed to
\beq
\label{e:nBILSR}
\min_{\z\in\bar{\mathcal{B}}}\|\bby-\bbR\z\|_2^2
\eeq
whose solution  is the BILS estimator of $\hbz$.

\section{Success Probabilities of the Babai estimators}\label{s:pba}

We consider the reduced box-constrained linear model \eqref{e:lmqr}.
The same analysis can be applied to
the transformed reduced linear model \eqref{e:nlmqr}.

The box-constrained Babai  estimator $\x^\sBB$ of $\hbx$ in \eqref{e:lmqr}, a suboptimal solution to \eqref{e:BILSR},
can be computed as follows:
\beq
\label{e:Babai}
\begin{split}
 c_{i}^\sBB&=(\ty_{i}-\sum_{j=i+1}^nr_{ij}x_j^\sBB)/r_{ii}, \\ \ \
 x_i^\sBB&=
\begin{cases}
\ell_i, & \mbox{ if }\   \lfloor c_i^\sBB\rceil\leq \ell_i\\
\lfloor c_i^\sBB\rceil, & \mbox{ if }\    \ell_i<\lfloor c_i^\sBB\rceil< u_i\\
u_i, & \mbox{ if }\    \lfloor c_i^\sBB\rceil \geq u_i
\end{cases}
\end{split}
\eeq
for $i=n, n-1, \ldots, 1$, where $\sum_{n+1}^n \cdot  =0$.
If we do not take the box constraint into account, we get the ordinary Babai estimator $\x^\sOB$:
\beq \label{e:BabaiO}
c_{i}^\sOB=(\ty_{i}-\sum_{j=i+1}^nr_{ij}x_j^\sOB)/r_{ii},\quad
 x_i^\sOB=
\lfloor c_i^\sOB \rceil  \\
\eeq
for $i=n, n-1, \ldots, 1$.

In the following, we  give  formulas  for the success probabilities  of $\x^\sBB$ and $\x^\sOB$.

\begin{theorem} \label{t:pmfpb}
Suppose that in  \eqref{e:lm0}  $\hbx$  is uniformly distributed over the constraint box $\mathcal{B}$, and $\hbx$ and ${\v}$ are independent.
Suppose that  \eqref{e:lm0} is transformed to  \eqref{e:lmqr} through the QR factorization \eqref{e:qr}.
Then the success probabilities of the box-constrained Babai estimator $\x^\sBB$ and the ordinary Babai estimator $\x^\sOB$,
which are respectively defined in \eqref{e:Babai} and \eqref{e:BabaiO}, are
\begin{align}
P^\sBB &\equiv \Pr(\x^\sBB=\hbx) \nonumber\\
&=\prod_{i=1}^n \Big[\frac{1}{u_i-\ell_i+1}+\frac{u_i-\ell_i}{u_i-\ell_i+1} \mbox{erf}\left(\frac{r_{ii}}{2\sqrt{2}\sigma}\right) \Big],  \label{e:pbbu} \\
 P^{\sOB} &\equiv \Pr(\x^\sOB=\hbx) = \prod_{i=1}^n  \mbox{erf}\left(\frac{r_{ii}}{2\sqrt{2}\sigma}\right),
\label{e:pob}
\end{align}
where  the error function  is
\[
\text{erf}(\zeta)=\frac{2}{\sqrt{\pi}}\int_{0}^{\zeta}\exp\big(-t^2\big)dt.
\]
\end{theorem}

{\bf Proof.} To simplify notation,  we denote
\begin{equation} \label{e:varphi}
\phi_\sigma(\zeta)= \mbox{erf}\left(\frac{\zeta}{2\sqrt{2}\sigma}\right)
\end{equation}
which will be used in this proof and other places.

Since the random vectors $\hbx$ and ${\v}$ in \eqref{e:lm0} are independent,
$\hbx$ and $\tilde{\v}$ in \eqref{e:lmqr} are also independent.
From \eqref{e:modelqr},
$$
\ty_i = r_{ii} \hx_i + \sum_{j=i+1}^n r_{ij}\hx_j + \tv_i, \ \   i=n,  n-1, \ldots, 1.
$$
Then from \eqref{e:Babai}, we obtain
\beq
c_i^\sBB = \hx_i + \sum_{j=i+1}^n \frac{r_{ij}}{r_{ii}} (\hx_j-x_j^\sBB) + \frac{\tv_i}{r_{ii}}, \ \   i=n,  n-1, \ldots, 1.
\label{e:ci}
\eeq
Therefore, if $x_{i+1}^\sBB=\hx_{i+1},\, \cdots, \,x_n^\sBB=\hx_n$ and $\hx_i$ is fixed, we have
$c_i^\sBB   \sim\mathcal{N}(\hx_{i},  \sigma^2/r_{ii}^2)$. Thus,
\beq \label{e:dis}
\frac{(c_i^\sBB -\hx_{i})r_{ii}}{\sqrt{2}\sigma}\sim\mathcal{N}\left(0,  \frac{1}{2}\right).
\eeq

To simplify notation, we denote  events
$$
E_i = (x_{i}^\sBB=\hx_{i},  \ldots, x_n^\sBB = \hx_n), \quad i=1,\ldots, n.
$$
Then, applying the chain rule of conditional probabilities yields
\beq\label{e:chain}
P^\sBB
 =\Pr(E_1) =\prod_{i=1}^{n}\Pr(x_i^\sBB=\hx_i|E_{i+1})
\eeq
where $E_{n+1}$ is the sample space $\Omega$ leading to $\Pr(x_n^\sBB=\hx_n|E_{n+1})=\Pr(x_n^\sBB=\hx_n)$.

Since events $\hx_i=\ell_i$, $\ell_i<\hx_i<u_i$ and $\hx_i=u_i$ are independent, by \eqref{e:Babai}, we have
\begin{align}
\label{e:bbprob0}
&\Pr(x_i^\sBB=\hx_{i} \,|\, E_{i+1})\nonumber\\
= & \Pr \left( (\hx_i=\ell_i, c_i^\sBB \leq \ell_i+1/2)   \,|\, E_{i+1}\right)  \nonumber\\
+&\Pr \left( (\ell_i<\hx_i<u_i, \hx_i-1/2\leq c_i^\sBB  < \hx_i + 1/2 )  \,|\, E_{i+1}\right)  \nonumber \\
+&\Pr \left((\hx_i=u_i, c_i^\sBB \geq u_i-1/2 ) \,|\, E_{i+1}\right).
\end{align}

In the following we will use this simple result: if $\bar{E}_1$, $\bar{E}_2$ and $\bar{E}_3$ are three events, and $\bar{E}_2$ and $\bar{E}_3$ are independent, then
\beq
\Pr((\bar{E}_1,\bar{E}_2)|\bar{E}_3)= \Pr(\bar{E}_1) \Pr(\bar{E}_2 |( \bar{E}_1, \bar{E}_3)) .
\label{e:probabc}
\eeq
This can easily be proved. In fact,
\begin{align*}
\Pr((\bar{E}_1,\bar{E}_2)|\bar{E}_3)&=\frac{\Pr(\bar{E}_1, \bar{E}_2,\bar{E}_3)}{\Pr(\bar{E}_3)}\\
&=\Pr(\bar{E}_1)\frac{\Pr(\bar{E}_1,\bar{E}_2, \bar{E}_3)}{\Pr(\bar{E}_1, \bar{E}_3)}\\
&= \Pr(\bar{E}_1) \Pr(\bar{E}_2 | (\bar{E}_1, \bar{E}_3)),
\end{align*}
where the second equality follows from the fact that $\bar{E}_1$ and $\bar{E}_3$ are independent.

Thus, by \eqref{e:bbprob0} and \eqref{e:probabc},  we obtain
\begin{align}
\label{e:bbprob}
&\Pr(x_i^\sBB=\hx_{i} \,|\, E_{i+1})\nonumber\\
=\, &  \Pr(\hx_i=\ell_i) \Pr \left( c_i^\sBB \leq \ell_i+1/2 \,|\, (\hx_i=\ell_i, E_{i+1})\right)   \nonumber  \\
  +&  \Pr(\ell_i<\hx_i<u_i) \nonumber\\
\times&\Pr \left(\hx_i-1/2\leq c_i^\sBB  < \hx_i + 1/2 \,|\, (\ell_i<\hx_i<u_i , E_{i+1})\right)  \nonumber \\
   +& \Pr( \hx_i=u_i) \Pr \left(c_i^\sBB \geq u_i-1/2  \,|\, ( \hx_i=u_i, E_{i+1})\right).
\end{align}

Since  $\hbx$  is uniformly distributed over the box $\mathcal{B}$,
for the first factors of the three terms on the right-hand side of \eqref{e:bbprob}, we have
\begin{align*}
\Pr(\hx_i=\ell_i)=\frac{1}{u_i-\ell_i+1},\\ \ \
\Pr(\ell_i<\hx_i<u_i) = \frac{u_i-\ell_i-1}{u_i-\ell_i+1}, \\\ \
\Pr(\hx_i=u_i)=\frac{1}{u_i-\ell_i+1}.
\end{align*}
By \eqref{e:varphi} and \eqref{e:dis}, for the second factors of these three terms, we have
\begin{align*}
& \Pr( c_i^\sBB \leq \ell_i+1/2 \,|\, (\hx_i=\ell_i, E_{i+1}))\\
 = & \Pr\left( \frac{(c_i^\sBB-\hx_i)r_{ii}}{\sqrt{2}\sigma} \leq \frac{r_{ii}}{2\sqrt{2}\sigma} \,\big|\, (\hx_i=\ell_i, E_{i+1})\right) \\
= & \frac{1}{\sqrt{\pi}}\int_{-\infty}^{\frac{r_{ii}}{2\sqrt{2}\sigma}}\exp\big(-t^2\big)dt
=  \frac{1}{2}\left[1+\phi_\sigma(r_{ii})\right],
\end{align*}
\begin{align*}
&\Pr(\hx_i-1/2\leq c_i^\sBB  < \hx_i + 1/2 \,|\, (\ell_i<\hx_i<u_i , E_{i+1}))\\
 = & \Pr\left(\Big| \frac{(c_i^\sBB-\hx_i)r_{ii}}{\sqrt{2}\sigma}\Big|
\leq \frac{r_{ii}}{2\sqrt{2}\sigma} \,\big|\, (\ell_i<\hx_i<u_i , E_{i+1})\right) \\
 =  & \frac{1}{\sqrt{\pi}}\int_{-\frac{r_{ii}}{2\sqrt{2}\sigma}}^{\frac{r_{ii}}{2\sqrt{2}\sigma}}\exp\big(-t^2\big)dt
=  \phi_\sigma(r_{ii}),
\end{align*}
\begin{align*}
&\Pr(c_i^\sBB \geq u_i-1/2  \,|\,  (\hx_i=u_i, E_{i+1}))\\
= & \Pr\left( \frac{(c_i^\sBB-\hx_i)r_{ii}}{\sqrt{2}\sigma} \geq -\frac{r_{ii}}{2\sqrt{2}\sigma} \,\big|\, (\hx_i=u_i, E_{i+1})\right) \\
= & \frac{1}{\sqrt{\pi}}\int^{\infty}_{-\frac{r_{ii}}{2\sqrt{2}\sigma}}\exp\big(-t^2\big)dt
=  \frac{1}{2}\left[1+\phi_\sigma(r_{ii})\right].
\end{align*}
Combining the  equalities above with \eqref{e:bbprob} yields
\begin{align*}
&\,\Pr(x_i^\sBB=\hx_{i} \,|\, E_{i+1})\nonumber \\
= &\, \frac{1}{2(u_i-\ell_i+1)} \left[1+\phi_\sigma(r_{ii})\right]
+ \frac{u_i-\ell_i-1}{u_i-\ell_i+1} \,\phi_\sigma(r_{ii})\nonumber \\
&+ \frac{1}{2(u_i-\ell_i+1)} \left[1+\phi_\sigma(r_{ii})\right]   \nonumber \\
 = & \,\frac{1}{u_i-\ell_i+1} + \frac{u_i-\ell_i}{u_i-\ell_i+1}\,\phi_\sigma(r_{ii})
\end{align*}
which, with \eqref{e:varphi} and \eqref{e:chain}, yields \eqref{e:pbbu}.

Now we consider the success probability of the ordinary Babai estimator $\x^\sOB$.
Everything in the first three paragraphs of this proof still holds if we replace each superscript BB by OB.
But we need to make more significant changes to the last two paragraphs.
We change \eqref{e:bbprob0} and \eqref{e:bbprob} as follows:
\begin{align*}
&\Pr(x_i^\sOB=\hx_{i} \,|\, E_{i+1})\\
=  & \Pr((\ell_i \leq \hx_i \leq u_i, \hx_i-1/2\leq c_i^\sOB  < \hx_i + 1/2 ) \,|\, E_{i+1})    \\
= & \Pr(\ell_i \leq \hx_i \leq u_i)\\
\times& \Pr(\hx_i-1/2\leq c_i^\sOB  < \hx_i + 1/2 \,|\, (\ell_i \leq \hx_i \leq u_i , E_{i+1})).
 \end{align*}
Here
\begin{align*}
& \Pr(\ell_i \leq \hx_i \leq u_i) =1, \\
& \Pr(\hx_i-1/2\leq c_i^\sOB  < \hx_i + 1/2 \,|\, (\ell_i \leq \hx_i \leq u_i , E_{i+1})) \\
&\quad= \phi_\sigma(r_{ii}).
\end{align*}
Thus
$$
\Pr(x_i^\sOB=\hx_{i} \,|\, E_{i+1})  = \phi_\sigma(r_{ii}).
$$
Then \eqref{e:pob} follows from  \eqref{e:varphi} and \eqref{e:chain} with each superscript BB replaced by OB.
 \ \ $\Box$
\medskip

From the proof of \eqref{e:pob}, we observe that the formula holds no matter what distribution of $\hbx$ is over the box $\mathcal{B}$.
Furthermore, the formula is identical to the one for the success probability of the ordinary Babai estimator $\x^\sOB$
when $\hbx$ in \eqref{e:lm0} is deterministic and is not subject to any box constraint; for more details, see \cite{ChaWX13}.

The following result shows the relationship between $P^\sBB$ and $P^\sOB$.
\begin{corollary} \label{c:bound}
Under the same assumption as in Theorem \ref{t:pmfpb},
\begin{align}
\label{e:BBLUBD}
P^{\sOB}  & < P^{\sBB},   \\
\lim_{\mbox{all} \,\;1\leq i\leq n, u_i-\ell_i \rightarrow \infty} P^\sBB & = P^\sOB.
 \label{e:pbblimit}
\end{align}
\end{corollary}

{\bf Proof.}
Note that
\[
\phi_\sigma(r_{ii})=\mbox{erf}(r_{ii} /(2\sqrt{2}\sigma)) < 1.
\]
 Thus
\begin{align*}
 \phi_\sigma(r_{ii})
&=  \frac{1}{u_i-\ell_i+1} \phi_\sigma(r_{ii}) + \frac{u_i-\ell_i}{u_i-\ell_i+1}\phi_\sigma(r_{ii})\\
& < \frac{1}{u_i-\ell_i+1} + \frac{u_i-\ell_i}{u_i-\ell_i+1}\phi_\sigma(r_{ii}).
\end{align*}
Then, by Theorem \ref{t:pmfpb}, we can conclude that  \eqref{e:BBLUBD} holds,
and we can also see \eqref{e:pbblimit} holds.
$\Box$

 \begin{corollary} \label{c:Babai1}
Under the same assumption as in Theorem \ref{t:pmfpb},
$P^\sBB$ and $P^\sOB$ increase  when $\sigma$ decreases and
$$
\lim_{\sigma\rightarrow 0}P^\sBB=\lim_{\sigma\rightarrow 0}P^\sOB=1.
$$
\end{corollary}

{\bf Proof.} For a given $\zeta$,
when $\sigma$ decreases $\mbox{erf}(r_{ii}/(2\sqrt{2}\sigma))$ increases
and $\lim\limits_{\sigma \rightarrow 0} \mbox{erf}(r_{ii}/(2\sqrt{2}\sigma)) =1$.
Then from Theorem \ref{t:pmfpb}, we immediately
see that the corollary holds.

%%%%%%%%%%%%%%%%%%%%%%%%%%%%%%%%%%%%%%%%%%%%%%%%%%%%%
\section{Effects of LLL-P, SQRD and V-BLAST on $P^\sBB$ }\label{s:proba}

Suppose that we perform the QR factorization  \eqref{e:permu} by using a column permutation strategy, such as LLL-P, SQRD or V-BLAST,
then we have the reduced box-constrained linear model \eqref{e:nlmqr}.
For \eqref{e:nlmqr} we can  define its corresponding Babai point $\z^\sBB$,
and use it as an estimator of $\hbz$, which is equal to $\P^{T}\hbx$, or equivalently
we use $\P\z^\sBB$ as an estimator of $\hbx$.

In this section, we will investigate how LLL-P, SQRD and V-BLAST column permutation strategies
affect the success probability $P^\sBB$ of the box-constrained Babai estimator.

\subsection{Effect of LLL-P on $P^\sBB$} \label{s:lll}

The LLL-P strategy involves a sequence of permutations of two consecutive columns of $\R$.
To investigate how LLL-P affects $P^\sBB$, we first look at one column permutation.
Suppose that $\delta\, r_{k-1,k-1}^2 > r^2_{k-1,k}+r^2_{kk}$ for some $k$ for the $\R$ in \eqref{e:lmqr}.
After the permutation of columns $k-1$ and $k$, $\R$ becomes $\bbR=\G_{k-1,k}^T\R \P_{k-1,k}$ (see \eqref{e:lllpermu}).
Then with the transformations given in \eqref{e:permtrans}, where $\bbQ=\G_{k-1,k}$ and $\P=\P_{k-1,k}$, \eqref{e:lmqr}  is transformed to \eqref{e:nlmqr}.
We will compare $\Pr(\x^\sBB=\hbx)$ and $\Pr(\z^\sBB=\hbz)$.

To prove our main results, we  need the following two lemmas.

\begin{lemma}
\label{l:funf}
Given $\alpha>0$, define
\begin{align}
\label{e:funf}
f(\zeta, \alpha) =&(1-2\zeta^2)\Big(1+\alpha\,\mbox{erf}(\zeta)\Big)
-\frac{2\alpha}{\sqrt{\pi}}\,\zeta\exp(-\zeta^2)
\end{align}
for $\zeta \geq 0$.
Then,  $f(\zeta, \alpha)$ is a strictly decreasing function of $\zeta$
and has  a unique zero $r(\alpha)$, i.e.,
\beq
f(r(\alpha), \alpha)=0.
\label{e:fga}
\eeq
When $\zeta > r(\alpha)$,  $f(\zeta, \alpha) <0$
and when $\zeta < r(\alpha)$, $f(\zeta,\alpha) > 0$.
Furthermore, $0< r(\alpha) <1/\sqrt{2}$, $r(\alpha)$ is a strictly decreasing function of $\alpha$,
and $\lim\limits_{\alpha\rightarrow \infty}r(\alpha)=0$.
\end{lemma}

{\bf Proof.} By some simple calculations, we obtain
\begin{align*}
\frac{\partial f(\zeta, \alpha)}{\partial \zeta}=-4\zeta\Big(1+\alpha\,\mbox{erf}(\zeta)\Big).
\end{align*}
Thus, for any $\zeta\geq 0$ and $\alpha > 0$, $\partial f(\zeta, \alpha) / \partial \zeta \leq 0$, where the equality holds if and only $\zeta=0$.
Therefore, $f(\zeta, \alpha)$ is a strictly  decreasing function of $\zeta$.

Note that $f(0,\alpha)=1>0$ and $f(1/\sqrt{2},\alpha)<0$ for $\alpha>0$,  by the implicit function theorem,
there exists a unique  $r(\alpha)$, which is continuously differentiable with respect to $\alpha$,
such that  \eqref{e:fga} holds
and $0 < r(\alpha) <1/\sqrt{2}$.
Since $f(\zeta, \alpha)$ is strictly  decreasing with respect to $\zeta$,  when $\zeta > r(\alpha)$,  $f(\zeta, \alpha) <0$
and when $\zeta < r(\alpha)$,  $f(\zeta, \alpha) >0$.

In the following, we show that $r(\alpha)$ is a strictly decreasing function of $\alpha$.
From \eqref{e:fga}, we have
\beq
\label{e:fga2}
\Big(1-2\,r^2(\alpha)\big)\big(1+\alpha\,\mbox{erf}(r(\alpha))\big)=\frac{2\alpha}{\sqrt{\pi}}\,r(\alpha)\exp\big(-r^2(\alpha)\big).
\eeq
Taking the derivative for both sides of \eqref{e:fga2} with respect to $\alpha$ yields
\begin{align*}
&-2\,r(\alpha)\,r'(\alpha)\big(1+\alpha\,\mbox{erf}(r(\alpha))\big)+\\
&\Big(1-2\,r^2(\alpha)\big)\left(\mbox{erf}(r(\alpha))
+\frac{2\alpha}{\sqrt{\pi}}\,r'(\alpha)\exp\big(-r^2(\alpha)\big)\right)\\
=&\frac{2}{\sqrt{\pi}}\,r(\alpha)\exp\big(-r^2(\alpha)\big)\\
&+\frac{2\alpha}{\sqrt{\pi}}\,r'(\alpha)\exp\big(-r^2(\alpha)\big)\Big(1-2\,r^2(\alpha)\big).
\end{align*}
Therefore,
\begin{align*}
&2\,r(\alpha)\,r'(\alpha)\big(1+\alpha\,\mbox{erf}(r(\alpha))\big)\\
=&(1-2\,r^2(\alpha))\,\mbox{erf}(r(\alpha))-\frac{2}{\sqrt{\pi}}\,r(\alpha)\exp\big(-r^2(\alpha)\big)\\
=&-\frac{1}{\alpha}(1-2\,r^2(\alpha)),
\end{align*}
where the latter equality follows from \eqref{e:fga2}.
Hence
\begin{align*}
r'(\alpha)=-\frac{(1-2\,r^2(\alpha))}{2\,\alpha\,r(\alpha)\big(1+\alpha\,\mbox{erf}(r(\alpha))\big)}<0.
\end{align*}

Finally, we show that $ \lim\limits_{\alpha\rightarrow \infty}r(\alpha)=0$.
Since $r(\alpha)$ is continuously differentiable with respect to $\alpha$  and $r(\alpha) >0 $ for $\alpha>0$, $\lim\limits_{\alpha\rightarrow \infty}r(\alpha)$ exists.
Let $\eta=\lim\limits_{\alpha\rightarrow \infty}r(\alpha)$, by the fact that $r(\alpha)$ is strictly decreasing with $\alpha$, we obtain that $0 \leq \eta \leq 1/\sqrt{2}$.

From \eqref{e:fga2}, we have
\begin{align*}
&\Big(1-2\,r^2(\alpha)\big)\mbox{erf}(r(\alpha))-\frac{2}{\sqrt{\pi}}\,r(\alpha)\exp\left( -r^2(\alpha)\right)\\
=&\frac{1-2\big(r(\alpha)\big)^2}{\alpha}.
\end{align*}
Then we take limits on both sides of the above equation as $\alpha\rightarrow \infty$, resulting in
\[
(1-2\,\eta^2)\,\mbox{erf}(\eta)-\frac{2}{\sqrt{\pi}}\,\eta\exp(-\eta^2)=0.
\]
Since $0 \leq \eta \leq 1/\sqrt{2}$, one can conclude from the above equation that  $\lim\limits_{\alpha\rightarrow+\infty}r(\alpha)=\eta=0$.
\ \ $\Box$

\medskip

\begin{remark}
Given $\alpha$, we can easily solve \eqref{e:fga} by a numerical method, e.g., the Newton method, to find $r(\alpha)$.
\end{remark}
 \medskip

\begin{lemma} \label{l:fung}
Given $\alpha, \beta> 0$, define
\begin{align}
\label{e:fung}
g(\zeta, \alpha, \beta) =\big(1+\alpha\,\mbox{erf}(\zeta)\big)\big(1+\alpha\,\mbox{erf}(\beta/\zeta)\big), \quad\zeta>0.
\end{align}
Then, when
\beq
\min\{\sqrt{\beta},\beta/r(\alpha)\} \leq \zeta< \max\{\sqrt{\beta},\beta/r(\alpha)\}
\label{e:zetab}
\eeq
where $r(\alpha)$ is defined in Lemma \ref{l:funf},
$g(\zeta, \alpha, \beta)$  is a strictly decreasing function of $\zeta$.
\end{lemma}

{\bf Proof.}
By the definition of $g$, we can easily obtain
\begin{align*}
\frac{\partial g(\zeta, \alpha, \beta) }{\partial \zeta}
=&\frac{2\alpha}{\sqrt{\pi}\zeta}\big(1+\alpha\,\mbox{erf}(\zeta)\big)
\big(1+\alpha\,\mbox{erf}(\beta/\zeta)\big)\\
&\times\left[h(\zeta, \alpha) - h(\beta/\zeta, \alpha) \right],
\end{align*}
where
\begin{align}
\label{e:funh}
h(\zeta, \alpha) =\frac{\zeta\exp (-\zeta^2 )}{1+\alpha\,\mbox{erf}(\zeta)}.
\end{align}
It is easy to see that in order to show the result,  we need only to show $h(\zeta, \alpha) < h(\beta/\zeta, \alpha)$
under the condition  \eqref{e:zetab} with $\zeta\neq \beta/\zeta$.

By some simple calculations and \eqref{e:funf}, we have
\begin{align}
\label{e:funhd}
\frac{\partial h(\zeta, \alpha)}{\partial \zeta}=
\frac{\exp(-\zeta^2)}{\Big(1+\alpha\,\mbox{erf}(\zeta)\Big)^2}\times f(\zeta, \alpha).
\end{align}
Now we assume that $\zeta$ satisfies \eqref{e:zetab} with $\zeta\neq \beta/\zeta$.
If  $\sqrt{\beta} < \beta/r(\alpha)$,   by \eqref{e:zetab}, we have $\zeta>\beta/\zeta >  r(\alpha)$, and
then from Lemma \ref{l:funf}, in this case, $f(\zeta,\alpha)<0$, thus  $\partial h(\zeta, \alpha) / \partial \zeta < 0$,
i.e.,   $h(\zeta, \alpha)$ is a strictly deceasing function of $\zeta$,
thus $h(\zeta, \alpha) < h(\beta/\zeta, \alpha)$.
If $\sqrt{\beta} > \beta/r(\alpha)$,   by \eqref{e:zetab}, we obtain $\zeta<\beta/\zeta <  r(\alpha)$, and
then from Lemma \ref{l:funf},  $f(\zeta,\alpha)>0$, thus  $\partial h(\zeta, \alpha)/ \partial \zeta > 0$, i.e.,
 $h(\zeta, \alpha)$ is a strictly increasing function of $\zeta$, thus again
$h(\zeta, \alpha) < h(\beta/\zeta, \alpha)$. \ \ $\Box$
\medskip

With the above lemmas, we can show how the success probability of the box-constrained Babai estimator changes
after two consecutive columns are swapped when the LLL-P strategy is applied. Specifically, we have the following theorem.

\begin{theorem} \label{t:enchancepb}
Suppose that in  \eqref{e:lm0}  the box $\mathcal{B}$ is a cube with edge length of  $d$,
$\hbx$  is uniformly distributed over  $\mathcal{B}$,  and $\hbx$ and ${\v}$ are independent.
Suppose that \eqref{e:lm0} is transformed to \eqref{e:lmqr} through the QR factorization
\eqref{e:qr} and  $\delta\, r_{k-1,k-1}^2 > r^2_{k-1,k}+r^2_{kk}$.
After the permutation of columns $k-1$ and $k$ of $\R$ and triangularization (see \eqref{e:lllpermu}), \eqref{e:lmqr} is transformed to  \eqref{e:nlmqr}.
\begin{enumerate}
\item If $r_{kk}\geq 2\sqrt{2}\,\sigma \,r(d)$, where $r(\cdot)$ is defined in Lemma \ref{l:funf},
then after the permutation, the success probability of the box-constrained Babai estimator increases, i.e.,
\beq
\Pr(\x^\sBB=\hbx) \leq \Pr(\z^\sBB=  \hbz).
\label{e:pbper3}
\eeq
\item If $r_{k-1,k-1}\leq 2\sqrt{2}\,\sigma \,r(d)$, then after the permutation,
the success probability of the box-constrained Babai estimator decreases,  i.e.,
\beq
\Pr(\x^\sBB=\hbx) \geq \Pr(\z^\sBB=  \hbz).
\label{e:pbper4}
\eeq
\end{enumerate}
Furthermore, the equality in each of \eqref{e:pbper3} and \eqref{e:pbper4} holds if and only if $r_{k-1,k}=0$.
\end{theorem}

{\bf Proof}.
When $r_{k-1,k}=0$, by Theorem \ref{t:pmfpb}, we see the equalities in  \eqref{e:pbper3} and \eqref{e:pbper4} hold.
In the following we assume $r_{k-1,k} \neq 0$ and show the strict inequalities in  \eqref{e:pbper3} and \eqref{e:pbper4} hold.

Define
\beq \label{e:km1k}
\beta \equiv  \frac{r_{k-1,k-1}}{2\sqrt{2}\sigma} \frac{r_{kk}}{2\sqrt{2}\sigma}
= \frac{\br_{k-1,k-1}}{2\sqrt{2}\sigma} \frac{\br_{kk}}{2\sqrt{2}\sigma}
\eeq
where for the second equality,  see \eqref{e:rbr}.
Using $\delta\, r_{k-1,k-1}^2 > r^2_{k-1,k}+r^2_{kk}$ and the equalities in \eqref{e:rbr},
we can easily verify that
\begin{align}
 \sqrt{\beta}
&\leq \max \Big\{\frac{\br_{k-1,k-1}}{2\sqrt{2}\sigma},\frac{\br_{kk}}{2\sqrt{2}\sigma}\Big\}\nonumber\\
&< \max \Big\{\frac{r_{k-1,k-1}}{2\sqrt{2}\sigma},\frac{r_{kk}}{2\sqrt{2}\sigma}\Big\}
=\frac{r_{k-1,k-1}}{2\sqrt{2}\sigma}\nonumber\\
&= \frac{\beta}{r_{kk}/(2\sqrt{2}\sigma)},  \label{in:maxmin}
\end{align}
\begin{align}
\frac{\beta}{r_{k-1,k-1}/(2\sqrt{2}\sigma)}
&=\frac{r_{kk}}{2\sqrt{2}\sigma}
= \min \Big\{\frac{r_{k-1,k-1}}{2\sqrt{2}\sigma},\frac{r_{kk}}{2\sqrt{2}\sigma}\Big\}\nonumber\\
&< \min\Big\{\frac{\br_{k-1,k-1}}{2\sqrt{2}\sigma},\frac{\br_{kk}}{2\sqrt{2}\sigma}\Big\}
\leq \sqrt{\beta}.  \label{in:maxmin1}
\end{align}

Now we prove part 1.
Note that after the permutation, $r_{k-1,k-1}$ and $r_{kk}$ change, but other diagonal entries of $\R$ do not change.
Then by Theorem \ref{t:pmfpb}, we can easily observe that \eqref{e:pbper3} is equivalent to
\begin{align}
\label{e:prxzk1}
 &\Big[\frac{1}{d+1}+\frac{d}{d+1}\mbox{erf}\left(\frac{r_{k-1,k-1}}{2\sqrt{2}\sigma}\right)\Big]
 \nonumber\\
&\times\Big[\frac{1}{d+1}+\frac{d}{d+1}\mbox{erf}\left(\frac{r_{kk}}{2\sqrt{2}\sigma}\right)\Big] \nonumber\\
\leq&
\Big[\frac{1}{d+1}+\frac{d}{d+1}\mbox{erf}\left(\frac{\br_{k-1,k-1}}{2\sqrt{2}\sigma}\right)\Big]
\nonumber\\
&\times\Big[\frac{1}{d+1}+\frac{d}{d+1}\mbox{erf}\left(\frac{\br_{kk}}{2\sqrt{2}\sigma}\right)\Big].
\end{align}
By \eqref{e:fung}, we can see that \eqref{e:prxzk1} is equivalent to
\begin{align}
\,&g\Big(\max \Big\{\frac{r_{k-1,k-1}}{2\sqrt{2}\sigma}, \frac{r_{kk}}{2\sqrt{2}\sigma}\Big\}, d, \beta \Big)\nonumber\\
\leq& g\Big(\max \Big\{\frac{\br_{k-1,k-1}}{2\sqrt{2}\sigma}, \frac{\br_{kk}}{2\sqrt{2}\sigma}\Big\}, d, \beta \Big).
\label{e:grand}
\end{align}
If  $r_{kk}\geq2\sqrt{2}\,\sigma \,r(d)$, then
the right-hand side of the last equality in \eqref{in:maxmin} satisfies
\beq
\frac{\beta}{r_{kk}/(2\sqrt{2}\sigma)} \leq \frac{\beta}{r(d)}.
\label{e:brkk}
\eeq
Then by combining \eqref{in:maxmin} and \eqref{e:brkk} and applying Lemma \ref{l:fung}
we can conclude that the strict inequality in \eqref{e:grand} holds.

The proof for part 2 is similar.
The inequality \eqref{e:pbper4} is equivalent to
\begin{align}
\,&g\Big(\min \Big\{\frac{r_{k-1,k-1}}{2\sqrt{2}\sigma}, \frac{r_{kk}}{2\sqrt{2}\sigma}\Big\}, d, \beta \Big)\nonumber\\
\geq& g\Big(\min \Big\{\frac{\br_{k-1,k-1}}{2\sqrt{2}\sigma}, \frac{\br_{kk}}{2\sqrt{2}\sigma}\Big\}, d, \beta \Big).
\label{e:grand1}
\end{align}
If $r_{k-1,k-1}\leq 2\sqrt{2}\,\sigma \,r(d)$, then
the left-hand side of the first equality in \eqref{in:maxmin1} satisfies
\beq
\frac{\beta}{r(d)} \leq \frac{\beta}{r_{k-1,k-1}/(2\sqrt{2}\sigma)}.
\label{e:brkk1}
\eeq
Then by combining \eqref{in:maxmin1} and \eqref{e:brkk1} and applying Lemma \ref{l:fung}
we can conclude that the strict inequality in \eqref{e:grand1} holds.
\ \ $\Box$
\medskip

We make a few remarks about  Theorem \ref{t:enchancepb}.

\begin{remark}
In the theorem, $\mathcal{B}$ is assumed to be a cube, not a more general box.
This restriction simplified the theoretical analysis. Furthermore, in practical applications, such as communications,
indeed  $\mathcal{B}$ is often a cube.
\end{remark}

\begin{remark}
After the permutation, the larger one of $r_{k-1,k-1}$ and $r_{kk}$ becomes smaller  (see  \eqref{in:maxmin})
and the smaller one  becomes larger
(see \eqref{in:maxmin1}),
so the gap between  $r_{k-1,k-1}$ and $r_{kk}$ becomes smaller.
This makes $P^\sBB$ increase under the condition $r_{kk}\geq 2\sqrt{2}\,\sigma \,r(d)$
or decrease   under the condition $r_{k-1,k-1}\leq 2\sqrt{2}\,\sigma \,r(d)$.
It is natural to ask for fixed $r_{k-1,k-1}$ and $r_{kk}$, when will $P^\sBB$ increase most or decrease most
after the permutation under the corresponding conditions?
From the proof we observe that  $P^\sBB$ will become maximal when the first inequality in \eqref{in:maxmin}
becomes an equality or minimal  when the last inequality in \eqref{in:maxmin1} becomes an equality
under the corresponding  conditions.
Either of the two equalities holds if and only if $\br_{k-1,k-1}=\br_{kk}$,
which is equivalent to $r_{k-1,k}^2+ r_{kk}^2 = r_{k-1,k-1}r_{kk}$ by \eqref{e:rbr}.
\end{remark}

\begin{remark}
The case where $r_{kk}< 2\sqrt{2}\,\sigma \,r(d) < r_{k-1,k-1}$ is not covered by the theorem.
For this case, $P^\sBB$ may increase or decrease after the permutation, for more details, see the simulations in Sec. \ref{s:sim}.
\end{remark}

\medskip

Based on Theorem \ref{t:enchancepb}, we can establish the following general result for
the LLL-P strategy.

\begin{theorem}\label{t:problll}
Suppose that in  \eqref{e:lm0}  the box $\mathcal{B}$ is a cube with edge length of  $d$,
$\hbx$  is uniformly distributed over  $\mathcal{B}$,  and $\hbx$ and ${\v}$ are independent.
Suppose that \eqref{e:lm0} is first transformed to  \eqref{e:lmqr} through the QR factorization \eqref{e:qr}
and then to  \eqref{e:nlmqr} through the QR factorization \eqref{e:permu} where the LLL-P strategy
is used for column permutations.
\ben
\item  If the diagonal entries of $\R$ in \eqref{e:lmqr} satisfies
\beq
\min_{1\leq i \leq n}r_{ii}\geq 2\sqrt{2}\,\sigma \,r(d),
\label{e:mincond}
\eeq
where $r(\cdot)$ is defined in Lemma \ref{l:funf}, then
\beq
\Pr(\x^\sBB=\hbx) \leq \Pr(\z^\sBB=\hbz).
\label{e:increase}
\eeq

\item  If the diagonal entries of $\R$ in \eqref{e:lmqr} satisfies
\beq
\max_{1\leq i \leq n}r_{ii}\leq 2\sqrt{2}\,\sigma \,r(d),
\label{e:maxcond}
\eeq
then
\beq
\Pr(\x^\sBB=\hbx) \geq \Pr(\z^\sBB=\hbz).
\label{e:decrease}
\eeq
\een
And the equalities in \eqref{e:increase} and \eqref{e:decrease} hold if and only if no column permutation occurs in the process or whenever two consecutive
columns, say $k-1$ and $k$, are permuted, $r_{k-1,k}=0$.
\end{theorem}

{\bf Proof.} It is easy to show that after each column permutation, the smaller one of the two diagonal entries of $\R$
involved in the permutation either keeps unchanged (the involved super-diagonal  entry is 0 in this case)
or strictly increases, while the larger one either keeps unchanged or strictly decreases
(see \eqref{in:maxmin} and \eqref{in:maxmin1}).
Thus, after each column permutation, the minimum of the diagonal entries of $\R$ either keeps unchanged or strictly increases
and the maximum either keeps unchanged or strictly decreases,
so  the diagonal entries of any upper triangular $\bbR$  produced after a column permutation  satisfies
$\min\limits_{1\leq i\leq n} r_{ii}\leq\br_{kk}\leq\max\limits_{1\leq i\leq n} r_{ii}$ for all $k=1,\ldots, n$.
Then the conclusion follows from Theorem \ref{t:enchancepb}. \ \ $\Box$

\medskip

We make some remarks about Theorem \ref{t:problll}.

\begin{remark}
 The quantity $r(d)$ is   involved in the conditions.
To get some idea about how large it is, we compute it for a few different $d=2^k-1$.
For $k=1,2,3,4,5$, the corresponding values of $r$ are  0.5939, 0.4926, 0.4042,  0.3286, 0.2653.
They are decreasing with $k$  as proved in Lemma \ref{l:funf}.
As $d \rightarrow \infty$, $r(d) \rightarrow 0$.
Thus, when $d$ is large enough, the condition   \eqref{e:mincond} will be satisfied.
By Corollary \ref{c:bound}, taking the limit as $d\rightarrow \infty$ on both sides of \eqref{e:increase}, we obtain
the following result proved in \cite{ChaWX13}:
$$
\Pr(\x^\sOB=\hbx) \leq \Pr(\z^\sOB=\hbz),
$$
i.e., LLL-P always increases the success probability of the ordinary Babai estimator.
\end{remark}

\begin{remark}
The two conditions \eqref{e:mincond} and \eqref{e:maxcond} also involve  the noise standard deviation $\sigma$.
When $\sigma$ is small, \eqref{e:mincond} is likely to hold, so applying LLL-P is likely to increase $P^\sBB$,
and when $\sigma$ is large, \eqref{e:maxcond} is likely to hold, so applying LLL-P is likely to decrease $P^\sBB$.
It is quite surprising that when $\sigma$ is large enough applying LLL-P will decrease $P^\sBB$.
Thus, before applying LLL-P, one needs to check the conditions  \eqref{e:mincond} and \eqref{e:maxcond}.
If  \eqref{e:mincond} holds, one has confidence to apply LLL-P.
If   \eqref{e:maxcond} holds, one should not apply it.
If both do not hold, i.e., $\min\limits_{1\leq i \leq n}r_{ii}< 2\sqrt{2}\,\sigma \,r(d) <  \max\limits_{1\leq i \leq n}r_{ii}$,
applying LLL-P may increase or decrease $P^\sBB$.

\end{remark}

\subsection{Effects of SQRD and V-BLAST on $P^\sBB$} \label{s:sqrdvblast}

SQRD and V-BLAST  have been used to find better ordinary and box-constrained Babai estimators in the literature.
It has been demonstrated in \cite{ChaWX13} that unlike LLL-P,  both SQRD and V-BLAST  may decrease the success probability
$P^\sOB$ of the ordinary Babai estimator when the parameter vector $\hbx$ is deterministic and not subject to any constraint.

We would like to know how SQRD and V-BLAST affect $P^\sBB$.
Unlike LLL-P, both SQRD and V-BLAST usually involve two non-consecutive columns permutations,
resulting in the changes of all diagonal entries between and including the two columns.
This makes it very difficult to analyze under what condition $\P^\sBB$ increases or decreases.
We will use numerical test results to show the effects of SQRD and V-BLAST on $P^\sBB$ with explanations.

In Theorem \ref{t:enchancepb} we showed that if the condition   \eqref{e:mincond} holds,
then applying LLL-P will increase $\P^\sBB$ and if \eqref{e:maxcond} holds, then applying LLL-P will decrease $\P^\sBB$.
%The following example shows there are not true for  SQRD and V-BLAST.
The following example shows that both SQRD and V-BLAST may decrease $\P^\sBB$ even if \eqref{e:mincond} holds,
and they may increase $\P^\sBB$ even if \eqref{e:maxcond} holds.

\bex \label{e:determin}
Let   $d=1$ and consider two matrices:
\beqnn
\R^{(1)} =\bmx    3.5  & 3 &   0\\
         0 &   1   & -1.5\\
         0    &     0  &  1
    \emx,
 \ \
\R^{(2)}=\bmx    1  & -1.5 &   1.5\\
         0 &   0.8   & -1\\
         0    &     0  &  0.42
    \emx.
\eeqnn

Applying SQRD, V-BLAST and LLL-P to $\R^{(1)}$ and $\R^{(2)}$, we obtain
\begin{align*}
\R^{(1)}_{\sss S}&=\bmx    1.8028  & -0.8321 &   0\\
         0 &   3.0509   & 3.4417\\
         0    &     0  &  0.6364
    \emx,\\
\R^{(1)}_{\sss V}=\R^{(1)}_{\sss L}&=\bmx    3.1623 &   3.3204  & -0.4743\\
         0 &    1.1068 &  1.4230\\
         0    &     0  &  1
    \emx,\\
\R^{(2)}_{\sss V}&=\bmx    1.7  & -1.7941 &   -0.8824\\
         0 &   0.4556   &  -0.1823\\
         0    &     0  &  0.4338
    \emx,\\
\R^{(2)}_{\sss S}&=\R^{(2)}_{\sss L}=\R^{(2)}.
\end{align*}

If $\sigma=0.2$, then it is easy to verify that for both $\R^{(1)}$ and $\R^{(2)}$, \eqref{e:mincond} holds
(note that $2\sqrt{2}\,r(d)= 2\sqrt{2}\,r(1)=1.6798$).
Simple calculations by using \eqref{e:pbbu} give
$$
P^\sBB(\R^{(1)})=0.9876, \ \
P^\sBB(\R^{(2)})=0.8286,
$$
and
$$
P^\sBB(\R^{(1)}_{\sss S})=0.9442, \ \  P^\sBB(\R^{(1)}_{\sss V})=P^\sBB(\R^{(1)}_{\sss L})=0.9910,
$$
$$
P^\sBB(\R^{(2)}_{\sss V})=0.7513, \ \ P^\sBB(\R^{(2)}_{\sss S})=P^\sBB(\R^{(2)}_{\sss L})=0.8286.
$$
Thus, SQRD decreases $P^\sBB$, while V-BLAST and LLL-P increase $P^\sBB$ for $\R^{(1)}$,
and V-BLAST decreases $P^\sBB$, while SQRD and LLL-P keep $P^\sBB$ unchanged for $\R^{(2)}$.

If $\sigma=2.2$, then it is easy to verify that for both $\R^{(1)}$ and $\R^{(2)}$, \eqref{e:maxcond} holds.
Simple calculations by using \eqref{e:pbbu} give
$$
P^\sBB(\R^{(1)})=0.2738, \ \
P^\sBB(\R^{(2)})=0.1816.
$$
Then
$$
P^\sBB(\R^{(1)}_{\sss S})=0.2777, \ \  P^\sBB(\R^{(1)}_{\sss V})=P^\sBB(\R^{(1)}_{\sss L})=0.2700,
$$
$$
P^\sBB(\R^{(2)}_{\sss V})=0.1898, \ \ P^\sBB(\R^{(2)}_{\sss S})=P^\sBB(\R^{(2)}_{\sss L})=0.1816.
$$
Thus,  SQRD increases $P^\sBB$, while V-BLAST and LLL-P decrease $P^\sBB$ for $\R^{(1)}$,
and  V-BLAST increases $P^\sBB$, while SQRD and LLL-P keep $P^\sBB$ unchanged for $\R^{(2)}$.
\eex
\medskip

Although Example \ref{e:determin} indicates that under the condition \eqref{e:mincond},
unlike LLL-P, both SQRD and V-BLAST may decrease  $P^\sBB$, often they increase $P^\sBB$.
This is the reason why SQRD and V-BLAST (especially the latter) have often been used to increase the accuracy of the Babai estimator in practice.
Example \ref{e:determin} also indicates that under the condition \eqref{e:maxcond}, unlike LLL-P,
both SQRD and V-BLAST may increase  $P^\sBB$, but often they decrease $P^\sBB$.
This is the opposite of what we commonly believe.
Later we will give numerical test results to show both phenomena.
In the following we give some explanations.

It is easy to show that like LLL-P, V-BLAST increases $\min_{1\leq i\leq} r_{ii}$  (not strictly) after each permutation
and  like LLL-P, SQRD decreases $\max_{1\leq i \leq n}r_{ii}$  (not strictly) after each permutation.
The relation between V-BLAST and SQRD can be found in \cite{ChaP07} and \cite{LinMG09}.
Thus if the condition \eqref{e:mincond} holds before applying V-BLAST, it will also hold after applying it;
and if the condition \eqref{e:maxcond} holds before applying SQRD, it will also hold after applying it.
Often applying V-BLAST decreases  $\max_{1\leq i \leq n}r_{ii}$  and  applying SQRD increases $\min_{1\leq i\leq} r_{ii}$
 (both may not be true sometimes, see Example \ref{e:determin}).
Thus often  the gaps between the large diagonal entries and the small ones of $\R$ decrease after applying SQRD or V-BLAST.
From the proof of Theorem \ref{t:enchancepb} we see reducing the gaps will likely increase $P^\sBB$ under  \eqref{e:mincond}
and decrease $P^\sBB$ under  \eqref{e:maxcond}.
Thus it is likely both SQRD and V-BLAST will increase $P^\sBB$ under \eqref{e:mincond}
and decrease it under \eqref{e:maxcond}.
We will give further explanations in the next subsection.

\subsection{A bound on $P^\sBB$}

In this subsection we give a bound on $P^\sBB$, which is an upper bound under \eqref{e:mincond}
and becomes a lower bound  under  \eqref{e:maxcond}.
This bound can help us to understand what a column permutation strategy should try to achieve.

 \begin{theorem}\label{t:LLLPbd}
Suppose that the assumptions in Theorem \ref{t:pmfpb} hold.
Let the box $\mathcal{B}$ in \eqref{e:box} be a cube with edge length of $d$
and  denote $\gamma=(\det(\R))^{1/n}$.
\begin{enumerate}
\item   If  the condition  \eqref{e:mincond} holds,
then
\beq
\label{e:LLLPUBD1}
\Pr(\x^\sBB=\hbx) \leq \left[\frac{1}{d+1}+\frac{d}{d+1}\mbox{erf}\Big(\frac{\gamma}{2\sqrt{2}\sigma}\Big)\right]^n.
\eeq

\item
If  the condition  \eqref{e:maxcond} holds,  then
\beq
\label{e:LLLPLBD1}
\Pr(\x^\sBB=\hbx) \geq \left[\frac{1}{d+1}+\frac{d}{d+1}\mbox{erf}\Big(\frac{\gamma}{2\sqrt{2}\sigma}\Big)\right]^n.
\eeq
\end{enumerate}
The equality in either  \eqref{e:LLLPUBD1} or \eqref{e:LLLPLBD1} holds if and only if
$r_{ii}=\gamma$ for $ i=1,\ldots, n$.

\end{theorem}

{\bf Proof.}
We prove only part 1.  Part 2 can be proved similarly.
Note that $\gamma^n= \Pi_{i=1}^n r_{ii}$.
Obviously, if $r_{ii}=\gamma$ for $i=1,\ldots, n$,
then by \eqref{e:pbbu}
the equality in   \eqref{e:LLLPUBD1}  holds. In the following we assume that there exist
$j$ and $k$ such that $r_{jj}\neq r_{kk}$,
we only need to show that the strict inequality  \eqref{e:LLLPUBD1} holds.

Denote $F(\zeta)=\ln(1+d\,\mbox{erf}\left(\frac{\exp(\zeta)}{2\sqrt{2}\sigma}\right))$,
$\eta_i=\ln(r_{ii})$ for $i=1,2,\ldots, n$ and $\eta=\frac{1}{n}\sum_{i=1}^n\eta_i$,
then by \eqref{e:pbbu}, \eqref{e:LLLPUBD1} is equivalent to
$$
\frac{1}{n}\sum_{i=1}^nF(\eta_i)< F(\eta).
$$
Since $\min_{1\leq i\leq} r_{ii} \geq 2\sqrt{2} \,\sigma \,r(d)$ and  $r_{jj}\neq r_{kk}$,
it suffices to show that $F(\zeta)$ is a strict concave function on
$(\ln(2\sqrt{2} \,\sigma \,r(d)),+\infty)$.
Therefore, we only need to show that $F''(\zeta)<0$ when $\zeta>\ln(2\sqrt{2} \,\sigma \,r(d))$.

To simplify  notation, denote $\xi=\exp(\zeta)/(2\sqrt{2}\sigma)$.  By some simple calculations, we obtain
$$
F'(\zeta)=\frac{d\,\xi\exp (-\xi^2)}{1+d\,\mbox{erf}(\xi)}=d\,h(\xi, d)
$$
where $h(\cdot,\cdot)$ is defined in  \eqref{e:funh}.
Then
$$
F''(\zeta) = d\, \xi\, (\partial h(\xi,d)/\partial \xi).
$$
By the proof of Lemma \ref{l:fung}, $\partial h(\xi,d)/\partial \xi<0$ when $\xi>r(d)$.
Thus, we can conclude that $F''(\zeta)< 0$ when $\zeta>\ln(2\sqrt{2} \,\sigma \,r(d))$, completing the proof.
\ \ $\Box$
\medskip

Now we make some remarks about Theorem \ref{t:LLLPbd}.

\begin{remark}\label{r:lllbd}
The quantity $\gamma$ is invariant with respect to column permutations, i.e., for $\R$ and $\bbR$  in  \eqref{e:permu},
we have the same $\gamma$ no matter what the permutation matrix $\P$ is.
Thus the bounds in \eqref{e:LLLPUBD1} and \eqref{e:LLLPLBD1}, which are actually the same quantity,
are invariant with respect to column permutations.
Although the condition  \eqref{e:mincond}  is variant with respect to column permutations,
if it holds before applying LLL-P or V-BLAST, it will hold afterwards,
since the minimum of the diagonal entries of $\bbR$ will not be smaller than that of $\R$
after applying LLL-P or V-BLAST.
Similarly, the condition  \eqref{e:maxcond} is also variant with respect to column permutations.
But if it holds before applying LLL-P or SQRD, it will hold afterwards,
since  the maximum of the diagonal entries of $\bbR$ will not be larger than that of $\R$
after applying LLL-P or  SQRD.
\end{remark}

\begin{remark}
The equalities in \eqref{e:LLLPUBD1} and \eqref{e:LLLPLBD1}  are reached if all the diagonal entries of $\R$ are identical.
This suggests that if the gaps between the larger entries and small entries become  smaller
after permutations, it is likely that $P^\sBB$ increases under  the condition  \eqref{e:mincond}
or decreases under the condition  \eqref{e:maxcond}.
As we know, the gap between the largest one and the smallest one decreases after applying LLL-P.
Numerical tests indicate usually this is also true for both V-BLAST and SQRD.
Thus  both V-BLAST and SQRD will likely bring $P^\sBB$ closer to the bound under the two opposite
conditions, respectively.
\end{remark}

\begin{remark}
When  $d \rightarrow \infty $, by Lemma \ref{l:funf}, $r(d) \rightarrow 0$,
thus the condition  in part 1 of Theorem \ref{t:LLLPbd} becomes $\max\limits_{1\leq i \leq n}r_{ii} \geq  0$, which always holds.
Taking the limit as $d \rightarrow \infty$ on both sides of \eqref{e:LLLPUBD1} and using Corollary \ref{c:bound},
we obtain
\beq
\label{e:LLLPUBD2}
\Pr(\x^\sOB=\hbx) \leq \left(\mbox{erf}(\gamma/(2\sqrt{2}\sigma))\right)^n.
\eeq
The above result was  obtained in \cite{Teu03} and  a simple proof was provided in \cite{ChaWX13}.
\end{remark}

\subsection{Numerical tests}\label{s:sim}

We have shown  that if \eqref{e:mincond} holds, then LLL-P increases $P^\sBB$  and \eqref{e:LLLPUBD1}
is an upper bound on $P^\sBB$; and if \eqref{e:maxcond} holds, then the LLL-P decreases $P^\sBB$ and \eqref{e:LLLPLBD1} is a lower bound on $P^\sBB$.
Example \ref{e:determin} in Sec.\ \ref{s:sqrdvblast} indicates that this conclusion does not always hold for SQRD and V-BLAST.
To further understand the effects of  LLL-P,  SQRD and V-BLAST on $P^\sBB$  and to see how
close they bring their corresponding $P^\sBB$  to the bounds given by \eqref{e:LLLPUBD1} and \eqref{e:LLLPLBD1},
we performed some numerical tests by \textsc{Matlab}.
For comparisons, we also performed tests for $P^\sOB$.

First we performed  tests for the following two cases:

\begin{itemize}
\item  Case 1. $\A$ is an $n\times n$ matrix whose entries are chosen independently and randomly
according to a zero mean Gaussian distribution with variance $1/2$.

\item  Case 2. $\A=\U\D\V^T$, $\U,\V$ are random orthogonal matrices obtained by the QR factorization of   matrices
whose entries are  chosen independently and andomly according to the standard Gaussian distribution
and $\D$ is an $n\times n$ diagonal matrix with $d_{ii}=10^{3(n/2-i)/(n-1)}$.
The condition number of  $\A$  is 1000.

\end{itemize}

In the tests for each case, we first chose $n=4$ and $\mathcal{B}=[0,1]^4$
and took different noise standard deviation $\sigma$ to test different situations according to the conditions \eqref{e:mincond} and
\eqref{e:maxcond} imposed  in Theorems \ref{t:problll} and \ref{t:LLLPbd}. The edge length $d$ of $\mathcal{B}$ is 1.
So in \eqref{e:mincond} and \eqref{e:maxcond},  $2\sqrt{2}\,r(d)= 2\sqrt{2}\,r(1)=1.6798$.
Details about choosing $\sigma$ will be given later.

We use $P^\sBB$,  $P^\sBB_\sL$, $P^\sBB_\sS$ and $P^\sBB_\sV$ respectively denote the success probability of the box-constrained Babai
estimator  corresponding to QR factorization (i.e., no permutations are involved), LLL-P, SQRD and V-BLAST,
and use $\mu^\sBB$ to denote  the right-hand side of  \eqref{e:LLLPUBD1} or \eqref{e:LLLPLBD1},
which is an upper bound if \eqref{e:mincond} holds and a lower bound if \eqref{e:maxcond} holds.
Similarly, $P^\sOB$,  $P^\sOB_{\sss L}$, $P^\sOB_{\sss S}$ and $P^\sOB_{\sss V}$ respectively denote the success probability of
the ordinary Babai estimator corresponding to QR factorization, LLL-P, SQRD and V-BLAST.
We use $\mu^\sOB$ to denote the right-hand side of \eqref{e:LLLPUBD2},
which is an upper bound on $P^\sOB$,  $P^\sOB_{\sss L}$, $P^\sOB_{\sss S}$ and $P^\sOB_{\sss V}$.
For each case, we performed 10 runs (notice that for each run we have different $\A$, $\hbx$ and $\v$ due to randomness)
and the results are displayed in Tables \ref{tb:1}-\ref{tb:6}.

In  Tables \ref{tb:1} and \ref{tb:4}, $\sigma=\sigma_1 \equiv \min\limits_{1\leq i \leq n} r_{ii}/1.8$.
It is easy to verify that the condition \eqref{e:mincond} holds.
This means that   $P^\sBB \leq P_\sL^\sBB$ by Theorem \ref{t:problll}
and $P^\sBB, P_\sL^\sBB, P_\sV^\sBB \leq \mu^\sBB$ by Theorem \ref{t:LLLPbd} and Remark \ref{r:lllbd}.
The numerical results given in Tables \ref{tb:1} and \ref{tb:4} are consistent with the theoretical results.
The numerical results also indicate that SQRD and V-BLAST  usually increase  (not strictly) $P^\sBB$,
although there is one exceptional case for SQRD in Table \ref{tb:4}.
We observe that the permutation strategies  increase $P^\sBB$ more significantly for Case 2 than for Case 1.
The reason is that $\A$ is more ill-conditioned for Case 2, resulting in larger gaps between the diagonal entries of $\R$,
which can  usually  be  reduced   more effectively by the permutation strategies.
We also observe that $P_\sS^\sBB \leq \mu^\sBB$ in both tables.
Although in theory the inequality  may not hold as we cannot guarantee
the condition \eqref{e:mincond} holds after applying SQRD, usually SQRD can make $\min_{1\leq i \leq n}r_{ii}$ larger.
Thus if  \eqref{e:mincond} holds before applying SQRD, it is likely that the condition still holds after applying it.
Thus it is likely that $P_\sS^\sBB \leq \mu^\sBB$ holds.

Tables \ref{tb:2} and \ref{tb:5} are opposite to  Tables \ref{tb:1} and \ref{tb:4}.
In both tables, $\sigma=\sigma_2 \equiv \max\limits_{1\leq i\leq n}r_{ii}/1.6$,  then the condition \eqref{e:maxcond} holds.
This means that   $P^\sBB \geq P_\sL^\sBB$ by Theorem \ref{t:problll}
and $P^\sBB, P_\sL^\sBB, P_\sS^\sBB \geq \mu^\sBB$ by Theorem \ref{t:LLLPbd} and Remark \ref{r:lllbd}.
The numerical results given in the two tables are consistent with the theoretical results.
The  results in the two tables also indicate that both SQRD and V-BLAST decrease (not strictly) $P^\sBB$,
although Example \ref{e:determin} shows that neither is always true under the condition \eqref{e:maxcond}.
We also observe that $P_\sV^\sBB \geq \mu^\sBB$ in both tables.
Although in theory the inequality  may not hold as we cannot guarantee
the condition \eqref{e:maxcond} holds after applying V-BLAST, usually V-BLAST can make $\max_{1\leq i \leq n}r_{ii}$ smaller.
Thus if  \eqref{e:maxcond} holds before applying V-BLAST, it is likely the condition still holds after applying it.
Thus it is likely $P_\sV^\sBB \geq \mu^\sBB$  holds.

In Tables \ref{tb:3} and \ref{tb:6},
\[
\sigma=\sigma_3 \equiv (0.3\max\limits_{1\leq i\leq n}r_{ii}+0.7\min\limits_{1\leq i\leq n}r_{ii})/1.68.
\]
In this case,
\[
\min\limits_{1\leq i \leq n} r_{ii} \leq \frac{1.68}{1.6798}2\sqrt{2}\sigma\, r(d) \leq \max\limits_{1\leq i\leq n}r_{ii},
\]
indicating that  \eqref{e:maxcond} does not hold and it is very likely that \eqref{e:mincond} does not hold either.
In theory we do not have results that cover this situation.
The numerical results in the two tables indicate all of the three permutation strategies can either increase or decrease  $P^\sBB$ strictly
and $\mu^\sBB$ can be larger or smaller than $P^\sBB$, $P_\sL^\sBB$, $P_\sS^\sBB$ and $P_\sV^\sBB$.
The reason we chose 0.3 and 0.7 rather than a more natural choice of 0.5 and 0.5 in defining $\sigma$ here
is that we may not be able to observe both increasing and decreasing phenomena due to limited runs.

Now we make comments on the success probability  of ordinary Babai points.
From Tables \ref{tb:1}-\ref{tb:6},
we observe that LLL-P  always  increases  (not strictly) $P^\sOB$,
and SQRD and V-BLAST almost always increases $P^\sOB$
(there is one exceptional case for SQRD in Table \ref{tb:4} and two exceptional cases for V-BLAST in Table \ref{tb:6}).
%After using these permutation strategies, $P^\sOB$ is often a good approximation to the upper bound $\mu^\sOB$.
%These observations are consistent with our finding given in \cite{ChaWX13}.
Thus the ordinary case is different from the box-constrained case.
We also observe $P^\sOB \leq P^\sBB$ for the same permutation strategies.
Sometimes the difference between the two is large (see Tables \ref{tb:5} and \ref{tb:6}).

\begin{table*}[ht]
\caption{Success probabilities and bounds for Case 1, $\sigma=\min_{1\leq i \leq n} r_{ii}/1.8$}
\centering
\begin{tabular}{|c||c|c|c|c|c||c|c|c|c|c|}
\hline
$\sigma$ & $P^\sBB$ & $P^\sBB_{\sss L}$ & $P^\sBB_{\sss S}$ &$P^\sBB_{\sss V}$ & $\mu^\sBB$ & $P^\sOB$ & $P^\sOB_{\sss L}$ & $P^\sOB_{\sss S}$ &$P^\sOB_{\sss V}$ & $\mu^\sOB$\\ \hline
0.0738 & 0.8159 & 1.0000 & 1.0000 & 1.0000 & 1.0000 & 0.6319 & 1.0000 & 1.0000& 1.0000 & 1.0000  \\ \hline
0.1537 & 0.7632 & 0.8423 & 0.8423 & 0.8423 & 0.9083 & 0.5503 & 0.6988 & 0.6988& 0.6988 & 0.8231  \\ \hline
0.1575 & 0.7938 & 0.9491 & 0.9491 & 0.9491 & 0.9698 & 0.5977 & 0.8998 & 0.8998& 0.8998 & 0.9403  \\ \hline
0.2170 & 0.7235 & 0.8577 & 0.8577 & 0.8577 & 0.8670 & 0.4893 & 0.7300 & 0.7300& 0.7300 & 0.7477  \\ \hline
0.1285 & 0.8133 & 0.8534 & 0.8534 & 0.8521 & 0.9882 & 0.6278 & 0.7070 & 0.7070& 0.7049 & 0.9766  \\ \hline
0.1676 & 0.6809 & 0.7529 & 0.7529 & 0.7529 & 0.8896 & 0.4255 & 0.5375 & 0.5375& 0.5375 & 0.7885  \\ \hline
0.3665 & 0.7039 & 0.7273 & 0.7273 & 0.7273 & 0.8004 & 0.4629 & 0.5093 & 0.5093& 0.5093 & 0.6324  \\ \hline
0.1968 & 0.6892 & 0.7320 & 0.7320 & 0.7385 & 0.8073 & 0.4420 & 0.5103 & 0.5103& 0.5270 & 0.6439  \\ \hline
0.3322 & 0.7087 & 0.7317 & 0.7317 & 0.7317 & 0.7665 & 0.4718 & 0.5156 & 0.5156& 0.5156 & 0.5765  \\ \hline
0.5221 & 0.4754 & 0.4754 & 0.4754 & 0.4754 & 0.4758 & 0.1899 & 0.1899 & 0.1899& 0.1899 & 0.1910  \\ \hline
\end{tabular}
\label{tb:1}
\end{table*}

\begin{table*}[htbp!]
\caption{Success probabilities and bounds for Case 2, $\sigma=\min_{1\leq i \leq n} r_{ii}/1.8$}
\centering
\begin{tabular}{|c||c|c|c|c|c||c|c|c|c|c|}
\hline
$\sigma$ & $P^\sBB$ & $P^\sBB_{\sss L}$ & $P^\sBB_{\sss S}$ &$P^\sBB_{\sss V}$ & $\mu^\sBB$ & $P^\sOB$ & $P^\sOB_{\sss L}$ & $P^\sOB_{\sss S}$ &$P^\sOB_{\sss V}$ & $\mu^\sOB$\\ \hline
0.0101 & 0.8155 & 0.9452 & 0.9354 & 0.9452 & 1.0000 & 0.6312 & 0.8905 & 0.8708& 0.8905 & 1.0000  \\ \hline
0.0130 & 0.7983 & 0.9839 & 0.9839 & 0.9839 & 1.0000 & 0.6045 & 0.9679 & 0.9679& 0.9679 & 1.0000  \\ \hline
0.0173 & 0.8159 & 0.9793 & 0.9793 & 0.9793 & 1.0000 & 0.6319 & 0.9586 & 0.9586& 0.9586 & 1.0000  \\ \hline
0.0066 & 0.8159 & 0.9913 & 0.9913 & 0.9967 & 1.0000 & 0.6319 & 0.9826 & 0.9826& 0.9933 & 1.0000  \\ \hline
0.0177 & 0.8106 & 0.9998 & 0.9998 & 0.9998 & 1.0000 & 0.6236 & 0.9997 & 0.9997& 0.9997 & 1.0000  \\ \hline
0.0060 & 0.8159 & 0.9841 & 0.9841 & 0.9998 & 1.0000 & 0.6319 & 0.9681 & 0.9681& 0.9996 & 1.0000  \\ \hline
0.0168 & 0.7833 & 0.8098 & 0.7625 & 0.8159 & 1.0000 & 0.5813 & 0.6224 & 0.5250& 0.6319 & 1.0000  \\ \hline
0.0150 & 0.8159 & 0.9999 & 0.9999 & 0.9999 & 1.0000 & 0.6319 & 0.9998 & 0.9998& 0.9998 & 1.0000  \\ \hline
0.0231 & 0.8159 & 0.9999 & 0.9999 & 0.9999 & 1.0000 & 0.6319 & 0.9999 & 0.9999& 0.9999 & 1.0000  \\ \hline
0.0211 & 0.7912 & 0.9696 & 0.9696 & 0.9892 & 1.0000 & 0.5935 & 0.9393 & 0.9393& 0.9784 & 1.0000  \\ \hline
\end{tabular}
\label{tb:4}
\end{table*}

\begin{table*}[htbp!]
\caption{Success probabilities and bounds for Case 1, $\sigma=\max(r_{ii})/1.6$}
\centering
\begin{tabular}{|c||c|c|c|c|c||c|c|c|c|c|}
\hline
$\sigma$& $P^\sBB$ & $P^\sBB_{\sss L}$ & $P^\sBB_{\sss S}$ &$P^\sBB_{\sss V}$ & $\mu^\sBB$ & $P^\sOB$ & $P^\sOB_{\sss L}$ & $P^\sOB_{\sss S}$ &$P^\sOB_{\sss V}$ & $\mu^\sOB$\\ \hline
1.1726 & 0.1557 & 0.1310 & 0.1310 & 0.1380 & 0.1121 & 0.0005 & 0.0006 & 0.0006& 0.0006 & 0.0006  \\ \hline
0.6432 & 0.2756 & 0.2756 & 0.2756 & 0.2756 & 0.2731 & 0.0387 & 0.0387 & 0.0387& 0.0387 & 0.0395  \\ \hline
0.5962 & 0.2915 & 0.2912 & 0.2912 & 0.2909 & 0.2900 & 0.0472 & 0.0473 & 0.0473& 0.0475 & 0.0478  \\ \hline
1.2435 & 0.1875 & 0.1632 & 0.1673 & 0.1632 & 0.1571 & 0.0040 & 0.0044 & 0.0044& 0.0044 & 0.0045  \\ \hline
0.8332 & 0.1873 & 0.1769 & 0.1769 & 0.1769 & 0.1750 & 0.0070 & 0.0074 & 0.0074& 0.0074 & 0.0074  \\ \hline
0.4875 & 0.2709 & 0.2709 & 0.2709 & 0.2709 & 0.2667 & 0.0356 & 0.0356 & 0.0356& 0.0356 & 0.0366  \\ \hline
0.9684 & 0.2769 & 0.2709 & 0.2709 & 0.2709 & 0.2688 & 0.0358 & 0.0369 & 0.0369& 0.0369 & 0.0375  \\ \hline
0.9971 & 0.1846 & 0.1665 & 0.1665 & 0.1665 & 0.1588 & 0.0043 & 0.0046 & 0.0046& 0.0046 & 0.0047  \\ \hline
1.2791 & 0.1501 & 0.1308 & 0.1308 & 0.1308 & 0.1294 & 0.0015 & 0.0016 & 0.0016& 0.0016 & 0.0016  \\ \hline
0.6327 & 0.2641 & 0.2564 & 0.2564 & 0.2564 & 0.2556 & 0.0301 & 0.0316 & 0.0316& 0.0316 & 0.0318  \\ \hline
\end{tabular}
\label{tb:2}
\end{table*}

\begin{table*}[htbp!]
\caption{Success probabilities and bounds for Case 2, $\sigma=\max(r_{ii})/1.6$}
\centering
\begin{tabular}{|c||c|c|c|c|c||c|c|c|c|c|}
\hline
$\sigma$ & $P^\sBB$ & $P^\sBB_{\sss L}$ & $P^\sBB_{\sss S}$ &$P^\sBB_{\sss V}$ & $\mu^\sBB$ & $P^\sOB$ & $P^\sOB_{\sss L}$ & $P^\sOB_{\sss S}$ &$P^\sOB_{\sss V}$ & $\mu^\sOB$\\ \hline
3.9438 & 0.1064 & 0.0947 & 0.0987 & 0.0947 & 0.0709 & 0.0000 & 0.0000 & 0.0000& 0.0000 & 0.0000  \\ \hline
2.4510 & 0.1173 & 0.1173 & 0.1173 & 0.1173 & 0.0764 & 0.0000 & 0.0000 & 0.0000& 0.0000 & 0.0000  \\ \hline
0.5790 & 0.1788 & 0.1640 & 0.1640 & 0.1640 & 0.1363 & 0.0019 & 0.0019 & 0.0019& 0.0019 & 0.0021  \\ \hline
5.3809 & 0.1011 & 0.0701 & 0.0701 & 0.0701 & 0.0686 & 0.0000 & 0.0000 & 0.0000& 0.0000 & 0.0000  \\ \hline
2.2574 & 0.1140 & 0.1023 & 0.0954 & 0.0954 & 0.0777 & 0.0000 & 0.0000 & 0.0000& 0.0000 & 0.0000  \\ \hline
3.7623 & 0.1099 & 0.0801 & 0.0801 & 0.0757 & 0.0713 & 0.0000 & 0.0000 & 0.0000& 0.0000 & 0.0000  \\ \hline
3.9225 & 0.1063 & 0.0834 & 0.0834 & 0.0834 & 0.0709 & 0.0000 & 0.0000 & 0.0000& 0.0000 & 0.0000  \\ \hline
1.3198 & 0.1153 & 0.1153 & 0.1153 & 0.1153 & 0.0900 & 0.0001 & 0.0001 & 0.0001& 0.0001 & 0.0001  \\ \hline
1.2416 & 0.1394 & 0.1108 & 0.1108 & 0.1108 & 0.0920 & 0.0001 & 0.0001 & 0.0001& 0.0001 & 0.0001  \\ \hline
0.8411 & 0.1719 & 0.1532 & 0.1532 & 0.1532 & 0.1090 & 0.0004 & 0.0004 & 0.0004& 0.0004 & 0.0005  \\ \hline
\end{tabular}
\label{tb:5}
\end{table*}

\begin{table*}[htbp]
\caption{Success probabilities and bounds for Case 1, $\sigma=(0.3\max(r_{ii})+0.7\min_{1\leq i \leq n} r_{ii})/1.68$}
\centering
\begin{tabular}{|c||c|c|c|c|c||c|c|c|c|c|}
\hline
$\sigma$ & $P^\sBB$ & $P^\sBB_{\sss L}$ & $P^\sBB_{\sss S}$ &$P^\sBB_{\sss V}$ & $\mu^\sBB$ & $P^\sOB$ & $P^\sOB_{\sss L}$ & $P^\sOB_{\sss S}$ &$P^\sOB_{\sss V}$ & $\mu^\sOB$\\ \hline
0.2848 & 0.4208 & 0.4336 & 0.4336 & 0.3184 & 0.2846 & 0.0154 & 0.0165 & 0.0165& 0.0252 & 0.0451  \\ \hline
0.6313 & 0.4720 & 0.4829 & 0.4829 & 0.4829 & 0.4863 & 0.1630 & 0.1932 & 0.1932& 0.1932 & 0.2017  \\ \hline
0.4328 & 0.4540 & 0.4599 & 0.4599 & 0.4599 & 0.4623 & 0.1517 & 0.1673 & 0.1673& 0.1673 & 0.1776  \\ \hline
0.6105 & 0.5054 & 0.5061 & 0.5061 & 0.5061 & 0.5092 & 0.2123 & 0.2161 & 0.2161& 0.2161 & 0.2259  \\ \hline
0.3306 & 0.4268 & 0.3807 & 0.3807 & 0.3805 & 0.3484 & 0.0321 & 0.0539 & 0.0539& 0.0539 & 0.0829  \\ \hline
0.2600 & 0.5055 & 0.5103 & 0.5103 & 0.5103 & 0.5252 & 0.1544 & 0.1868 & 0.1868& 0.1868 & 0.2437  \\ \hline
0.4743 & 0.4235 & 0.4283 & 0.4283 & 0.4283 & 0.4259 & 0.0631 & 0.1225 & 0.1225& 0.1225 & 0.1437  \\ \hline
0.5878 & 0.4104 & 0.4161 & 0.4161 & 0.4161 & 0.4170 & 0.1159 & 0.1304 & 0.1304& 0.1304 & 0.1359  \\ \hline
0.3977 & 0.4429 & 0.4431 & 0.4431 & 0.4431 & 0.4477 & 0.1477 & 0.1479 & 0.1479& 0.1479 & 0.1636  \\ \hline
0.6273 & 0.4684 & 0.4684 & 0.4684 & 0.4684 & 0.4696 & 0.1792 & 0.1792 & 0.1792& 0.1792 & 0.1848  \\ \hline
\end{tabular}
\label{tb:3}
\end{table*}

\begin{table*}[htbp]
\caption{Success probabilities and bounds for Case 2, $\sigma=(0.3\max(r_{ii})+0.7\min_{1\leq i \leq n} r_{ii})/1.68$}
\centering
\begin{tabular}{|c||c|c|c|c|c||c|c|c|c|c|}
\hline
$\sigma$ & $P^\sBB$ & $P^\sBB_{\sss L}$ & $P^\sBB_{\sss S}$ &$P^\sBB_{\sss V}$ & $\mu^\sBB$ & $P^\sOB$ & $P^\sOB_{\sss L}$ & $P^\sOB_{\sss S}$ &$P^\sOB_{\sss V}$ & $\mu^\sOB$\\ \hline
1.0377 & 0.1608 & 0.1324 & 0.1324 & 0.1625 & 0.0987 & 0.0001 & 0.0002 & 0.0002& 0.0002 & 0.0002  \\ \hline
0.3648 & 0.2774 & 0.2774 & 0.2774 & 0.2405 & 0.1987 & 0.0034 & 0.0034 & 0.0034& 0.0025 & 0.0126  \\ \hline
0.7603 & 0.1681 & 0.1758 & 0.1758 & 0.1758 & 0.1150 & 0.0003 & 0.0005 & 0.0005& 0.0005 & 0.0007  \\ \hline
0.8769 & 0.1835 & 0.2062 & 0.1713 & 0.2062 & 0.1067 & 0.0002 & 0.0003 & 0.0004& 0.0003 & 0.0004  \\ \hline
0.4708 & 0.2794 & 0.2352 & 0.2352 & 0.2352 & 0.1590 & 0.0010 & 0.0030 & 0.0030& 0.0030 & 0.0048  \\ \hline
1.1983 & 0.1572 & 0.1319 & 0.1319 & 0.1319 & 0.0932 & 0.0001 & 0.0001 & 0.0001& 0.0001 & 0.0001  \\ \hline
1.0001 & 0.1758 & 0.1596 & 0.1596 & 0.1464 & 0.1003 & 0.0001 & 0.0002 & 0.0002& 0.0001 & 0.0002  \\ \hline
0.8523 & 0.1671 & 0.1733 & 0.1733 & 0.1715 & 0.1082 & 0.0002 & 0.0003 & 0.0003& 0.0003 & 0.0005  \\ \hline
0.2128 & 0.3478 & 0.3478 & 0.3728 & 0.3478 & 0.3539 & 0.0599 & 0.0599 & 0.0711& 0.0599 & 0.0866  \\ \hline
0.3956 & 0.2188 & 0.2117 & 0.2117 & 0.1973 & 0.1844 & 0.0047 & 0.0047 & 0.0047& 0.0034 & 0.0093  \\ \hline
\end{tabular}
\label{tb:6}
\end{table*}

Each of Tables \ref{tb:1}-\ref{tb:6} displays the results for only 10 runs due to space limitation.
To make up for this shortcoming,  we give Tables \ref{tb:ICD1} and \ref{tb:ICD2}, which display
some statistics for 1000 runs on the data generated exactly the same way as the data for the 10 runs.
Specifically these two tables display the number of runs,  in which
$P^\sBB$ ($P^\sOB$) strictly increases, keeps  unchanged and strictly decreases
after each of the three permutation strategies is applied for Case 1 and Case 2, respectively.
In the two tables,   $\sigma_1$, $\sigma_2$ and $\sigma_3$ are defined in the same as those  used in Tables \ref{tb:1}--\ref{tb:6}.

\begin{table*}[tbp!]
\caption{Number of runs out of 1000 in which $P^\sBB$ and $P^\sOB$ changes for Case 1}
\vspace*{-2mm}
\centering
\begin{tabular}{|c||c||c|c|c||c|c|c|}
 \hline
 \multicolumn{2}{|c||}{}  &  \multicolumn{3}{c||}{$P^\sBB$} & \multicolumn{3}{c|}{$P^\sOB$} \\ \hline
$\sigma$  &  \backslashbox{Result}{Strategy} & LLL-P &  SQRD  &  V-BLAST  &   LLL-P  &  SQRD  &  V-BLAST  \\ \hline
 & Strict increase  & 933  & 928 &  951 &  933  & 922 &  953 \\  % \cline{2-8}
$\sigma_1$ & No change  & 67  &  47 &   42  &  67  &  47  &  42\\  % \cline{2-8}
  &  Strict decrease  &   0  &  25 &    7 &    0 &   31  &   5 \\  \hline
 &  Strict  increase  &0  &  25 &    6  & 942 &  947 &  950 \\  %\cline{2-8}
$\sigma_2$ & No  change  & 58  &  40 &   37 &   58&    40 &   37\\  % \cline{2-8}
 &  Strict decrease   &  942 &   935 &  957 &    0  &  13  &  13\\ \hline
&  Strict increase  & 781 &  797 &  740 &  942 &  945 &  952 \\  %\cline{2-8}
$\sigma_3$ & No  change  &  58 &   40 &   37 &   58  &  40 &   37 \\  % \cline{2-8}
 &  Strict decrease  &161  & 163 &  223   &  0  &  15  &  11\\ \hline
\end{tabular}
\label{tb:ICD1}
\end{table*}

\begin{table*}[tbp!]
\caption{Number of runs out of 1000 in which $P^\sBB$ and $P^\sOB$ changes  for Case 2}
\vspace*{-2mm}
\centering
\begin{tabular}{|c||c||c|c|c||c|c|c|}
 \hline
 \multicolumn{2}{|c||}{}  &  \multicolumn{3}{c||}{$P^\sBB$} & \multicolumn{3}{c|}{$P^\sOB$} \\ \hline
$\sigma$   &  \backslashbox{Result }{Strategy} &  LLL-P  &  SQRD  &  V-BLAST  &   LLL-P  &  SQRD  &  V-BLAST  \\ \hline
&  Strict increase  &    858 &  803 &  938  & 858 &  800 &  938 \\  % \cline{2-8}
$\sigma_1$  & No  change  & 142  &  76 &   56 &  142 &   76 &   56\\  % \cline{2-8}
  &  Strict decrease  &  0  & 121  &   6 &    0  & 124 &    6\\  \hline
 &  Strict increase  & 0  &  23 &   69 &  906 &  944 &  831 \\  %\cline{2-8}
$\sigma_2$ & No change  &  94 &   46 &   48 &   94 &   46 &   48\\  % \cline{2-8}
 &  Strict decrease   &   906  & 931 &  883 &    0  &  10 &  121\\ \hline
&  Strict increase  & 134 &  189 &   97  & 906 &  943  & 840 \\  %\cline{2-8}
$\sigma_3$ & No  change  &  94  &  46 &   48  &  94 &   46  &  48 \\  % \cline{2-8}
 &  Strict decrease  &772  & 765&   855 &    0  &  11 &  112\\ \hline
\end{tabular}
\label{tb:ICD2}
\end{table*}

From Tables \ref{tb:ICD1} and \ref{tb:ICD2},  we can see that often these permutation strategies increase
or decrease $P^\sBB$  for the same data.
The numerical results  given in all the tables suggest that if the condition \eqref{e:mincond} holds,
we should have confidence to use any of these permutation strategies;
and if the condition \eqref{e:maxcond} holds we should not use any of them.

Tables \ref{tb:ICD1} and \ref{tb:ICD2} do  not show which permutation  strategy   increases $P^\sBB$ most for small $\sigma$.
The information on this given in Tables \ref{tb:1}-\ref{tb:6} are limited.
In the following we give more test results to investigate this.

As the main application of this research is in digit communications,
we used the MIMO model in the new tests.
For a fixed dimension, a fixed type of QAM and a fixed $E_b/N_0$,
we randomly generated 200 complex channel matrices
whose entries independently and identically follow the standard complex normal distribution,
and  for each generated channel matrix, we randomly generated 500 pairs of complex signal vector
(whose entries  are uniformly distributed according to the QAM constellation)
and complex  noise vector (whose entries are independently and identically normally distributed),
resulting in 10000 instances of a complex linear model.
Each complex instance was then transformed to an instance of the real linear model  \eqref{e:lm0}.

Unlike the previous tests, we compare the {\em experimental} error probabilities of the box-constrained Babai estimators
(i.e., the ratio of the number of runs that the Babai point is not equal to the true parameter vector $\hbx$ to 10000)
corresponding to QR, LLL-P, SQRD and V-BLAST, and the {\em theoretical} bound on the error probability of a Babai estimator
(i.e., the difference between 1 and the bound on its success probability (see \eqref{e:LLLPUBD1})).

\begin{figure}[!htbp]
\centering
\includegraphics[width=3.2in]{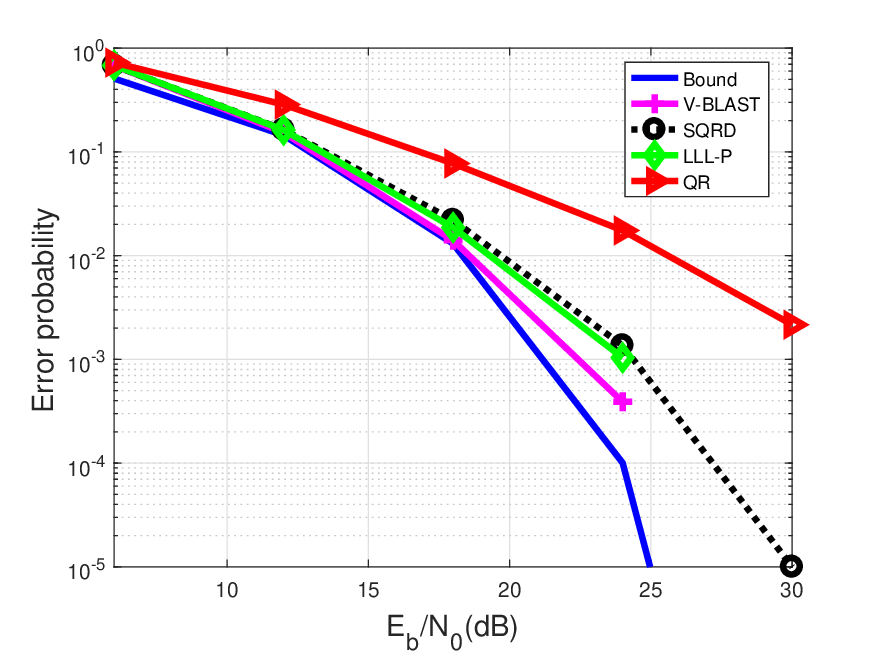}
\caption{Error probability of Babai point and bound versus $E_b/N_0=5\!:\!5\!:\!30$ for the $4\times4$ MIMO system with 16-QAM}
\label{fig:416}
\end{figure}

\begin{figure}[!htbp]
\centering
\includegraphics[width=3.2in]{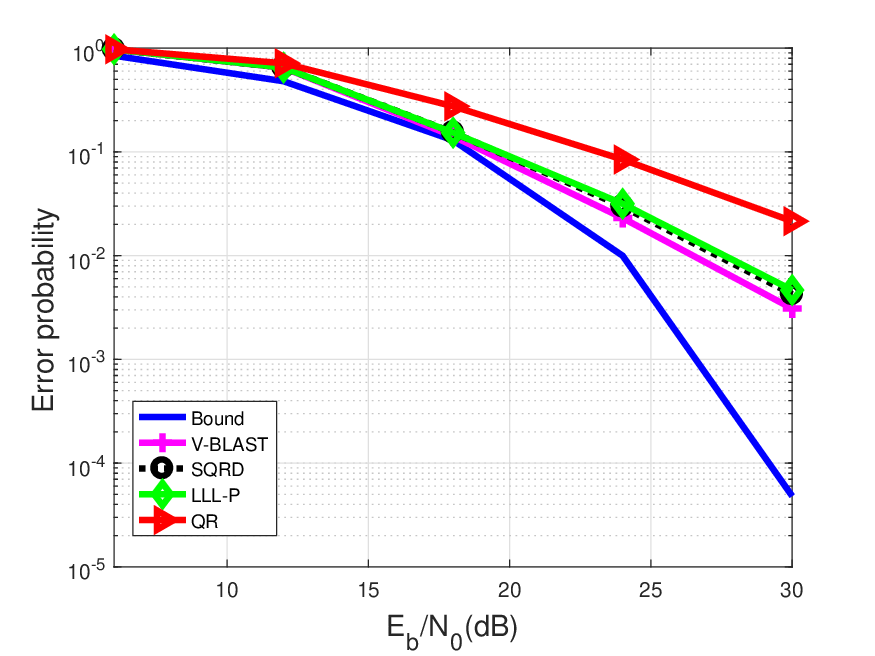}
\caption{Error probability of Babai point and bound versus $E_b/N_0=5\!:\!5\!:\!30$ for the $4\times4$ MIMO system with 64-QAM}
\label{fig:464}
\end{figure}

\begin{figure}[!htbp]
\centering
\includegraphics[width=3.2in]{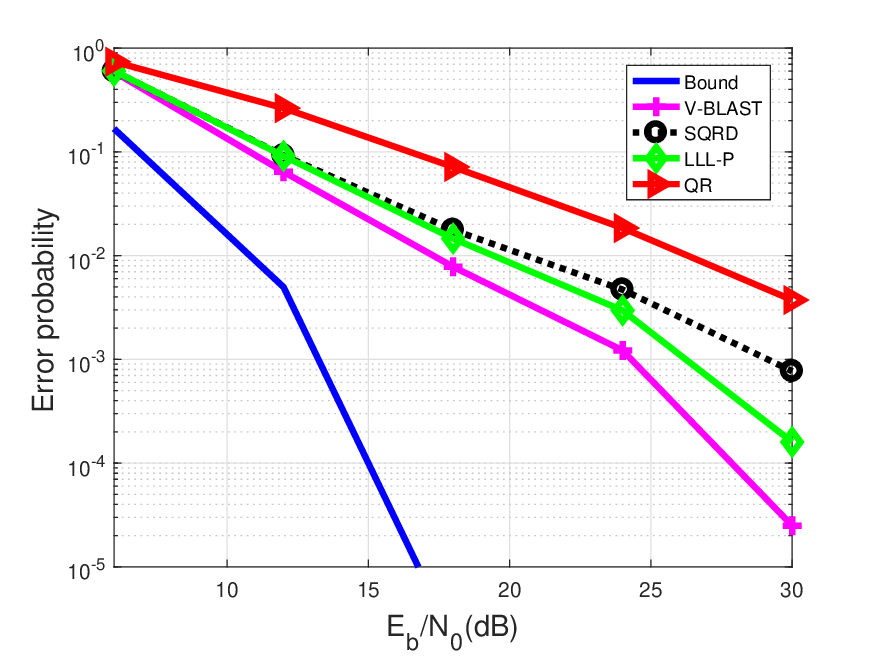}
\caption{Error probability of Babai point and bound versus $E_b/N_0=5\!:\!5\!:\!30$ for the $8\times8$ MIMO system with 16-QAM}
\label{fig:816}
\end{figure}

\begin{figure}[!htbp]
\centering
\includegraphics[width=3.2in]{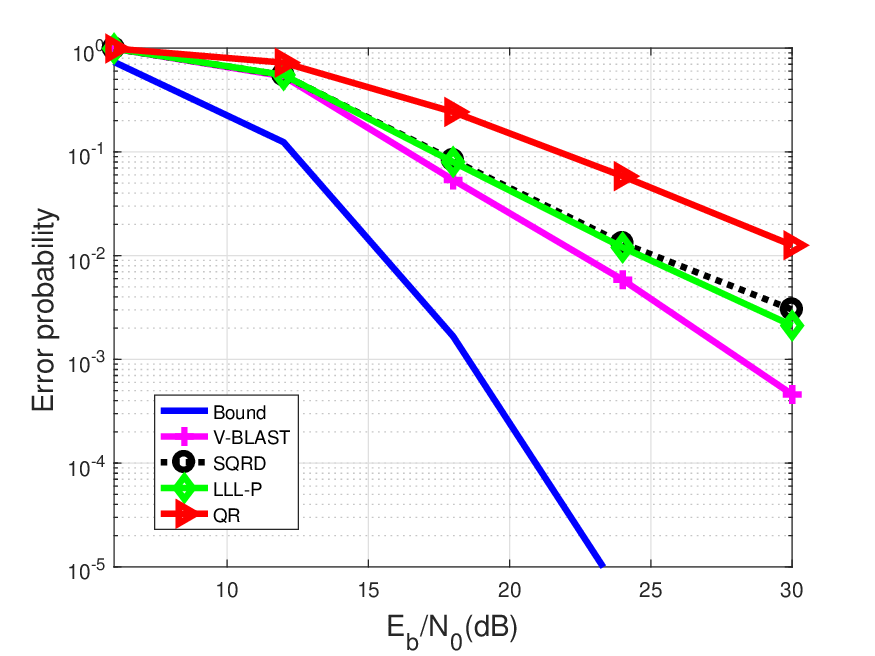}
\caption{Error probability of Babai point and bound versus $E_b/N_0=5\!:\!5\!:\!30$ for the $8\times8$ MIMO system with 64-QAM}
\label{fig:864}
\end{figure}

\begin{figure}[!htbp]
\centering
\includegraphics[width=3.2in]{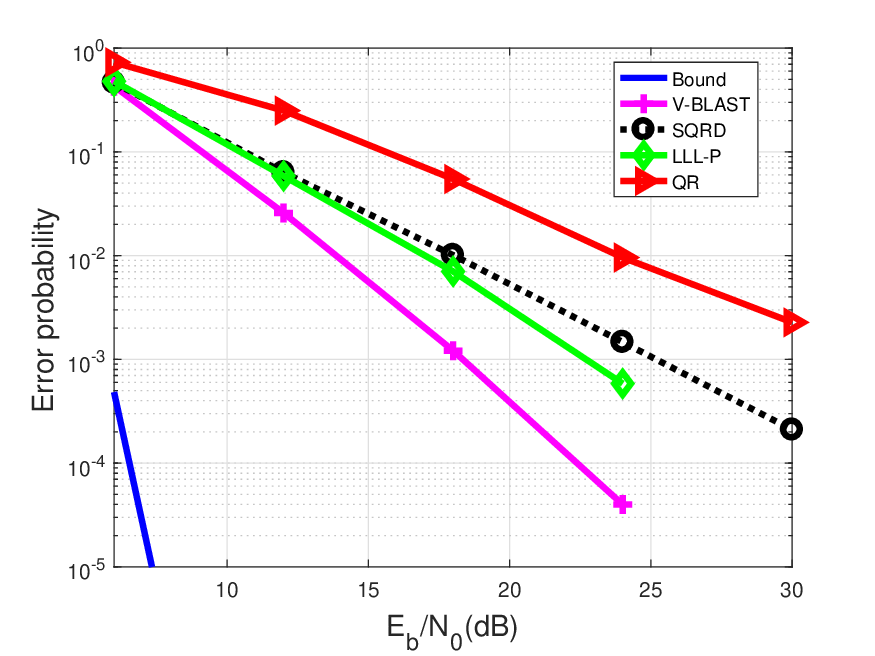}
\caption{Error probability of Babai point and bound versus $E_b/N_0=5\!:\!5\!:\!30$ for the $16\times16$ MIMO system with 16-QAM}
\label{fig:1616}
\end{figure}

\begin{figure}[!htbp]
\centering
\includegraphics[width=3.2in]{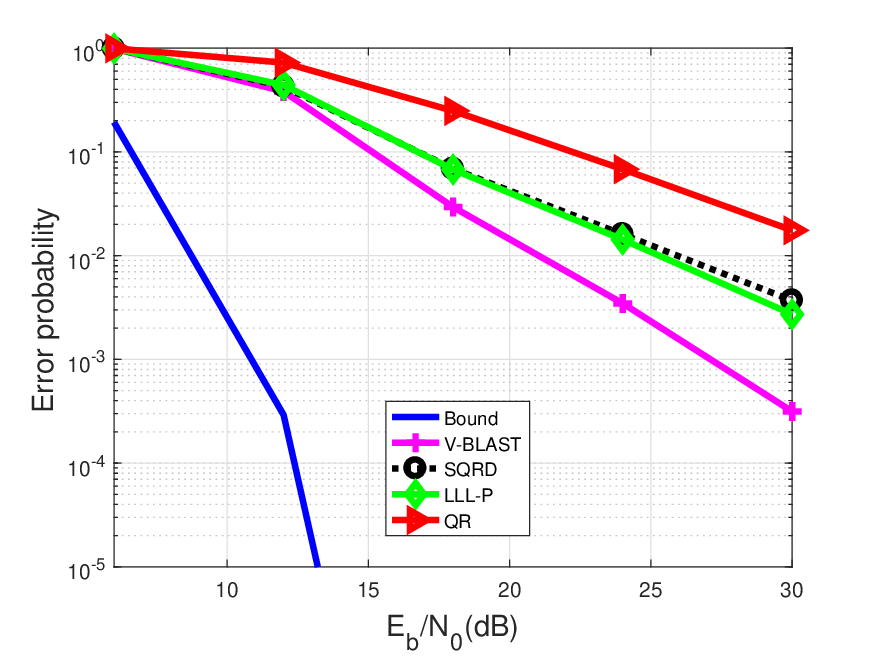}
\caption{Error probability of Babai point and bound versus $E_b/N_0=5\!:\!5\!:\!30$ for the $16\times16$ MIMO system with 64-QAM}
\label{fig:1664}
\end{figure}

Figures \ref{fig:416} and \ref{fig:464} respectively display the  experimental error probability corresponding to the QR factorization,
and the three permutation strategies, and the average  theoretical bound over the 10000 runs
versus $E_b/N_0=5\!:\!5\!:\!30$ for the $4\times4$ MIMO system with 16-QAM and 64-QAM.
Similarly, Figures \ref{fig:816} and \ref{fig:864} respectively show the corresponding results
 for the $8\times8$ MIMO system with 16-QAM and 64-QAM.
And Figures \ref{fig:1616} and \ref{fig:1664} show the corresponding results
for the $16\times16$ MIMO system with 16-QAM and 64-QAM, respectively.

From Figures \ref{fig:416}-\ref{fig:1664}, we can see that on average all of the three column
permutation strategies decrease the error probability of the Babai point and
the error bound is a lower bound (this is because \eqref{e:mincond} usually holds, which ensures \eqref{e:LLLPUBD1}).
These Figures also show that the effect of V-BLAST is much more significant
than that of  LLL-P and  SQRD, which have more or less the same performance.
This phenomenon is  similar to that for $P^\sOB$, as shown in \cite{ChaWX13}.

%%%%%%%%%%%%%%%%%%%%%%%%%%%%%%%%%%%%%%%%%%%%%%%%%

\section{On the Conjecture proposed in \cite{MaHFL09} }\label{s:conjecture}

In \cite{MaHFL09}, a conjecture was made on the ordinary Babai estimator,
based on which a stopping criterion was then proposed for the sphere decoding search process
for solving the BILS problem \eqref{e:BILS}.
In this section, we  first introduce this conjecture, then give an example to show that this conjecture
may not hold in general, and finally  we show that the conjecture holds under some conditions.

The problem considered in  \cite{MaHFL09} is to estimate
the  integer parameter vector $\hbx$ in the box-constrained linear model \eqref{e:lm0}.
The method proposed in \cite{MaHFL09} first ignores the box constraint \eqref{e:box}.
Instead of using the column permutations in \eqref{e:permu}, it performs the LLL reduction:
\beq \label{e:LLLC}
\bbQ^T\R\Z=\bbR
\eeq
where $\bbQ$ is  orthogonal, $\Z$ is  unimodular  (i.e, $\Z\in \mathbb{Z}^{n\times n}$ and $\det(\Z)=\pm1$ ) and
the upper triangular $\bbR$ is LLL reduced, i.e., it satisfies the Lov\'{a}sz condition \eqref{e:lovasz} and the size-reduce condition:
\[
|\br_{ik}|\leq\frac{1}{2}\bar{r}_{ii}, \quad k=i+1, i+2, \ldots, n, \quad i=1,2,\ldots,n-1.
\]
Then, with $\bar{\y}=\bar{\Q}^T\tilde{\y}$, $\bar{\v}=\bar{\Q}^T\tilde{\v}$ and $\hat{\z}=\Z^{-1}\hat{\x}$,
the ordinary linear model \eqref{e:modelqr} %and \eqref{e:OILSR}
becomes
\beq \label{e:modelC}
\bar{\y}=\bbR\hat{\z}+\bar{\v}, \ \ \bar{\v}\sim\mathcal{N}(\boldsymbol{0},\sigma^2 \I).
\eeq
For the reduced model, one can find its ordinary Babai estimator $\z^\sOB$ (c.f.\ \eqref{e:BabaiO}):
\beq \label{e:BabaiC}
 c_{i}^\sOB=(\bar{y}_{i}-\sum_{j=i+1}^n\bar{r}_{ij}z_j^\sOB)/\bar{r}_{ii}, \ \ z_i^\sOB=\lfloor c_i^\sOB\rceil
\eeq
for $i=n, n-1, \ldots, 1$, where $\sum_{j=n+1}^n\cdot=0$.
Define $\bar{\x}=\Z\z^\sOB$.
In \cite{MaHFL09}, $\bar{\x}$ is used to estimate the true parameter vector $\hbx$.
If  $\bar{\x}\neq \hat{\x}$, then a vector error (VE) is said to have occurred.
Note that $\bar{\x}$ may be outside the constraint box $\mathcal{B}$ in \eqref{e:box}.
If $\bar{\x}\in \mathcal{B}$, then $\bar{\x}$ is called a  valid vector, otherwise, i.e., $\bar{\x} \notin \mathcal{B}$,
$\bar{\x}$ is called an invalid vector.
The conjecture proposed in \cite{MaHFL09} is: a VE is most likely to occur
if $\bar{\x}$ is invalid; conversely, if $\bar{\x}$ is valid, there is little chance that the vector is in error.

From the definition of VE, if $\bar{\x}$ is invalid, then VE must occur. So in the following, we will only consider the second part of the conjecture,
i.e.,   $\Pr(\bar{\x}\neq\hbx|\bar{\x}\in \mathcal{B}) \approx 0$.

\subsection{The conjecture does not always hold }

In this subsection, we first show that $\Pr(\bar{\x}\neq\hbx|\bar{\x}\in \mathcal{B})$ can be very close to 1,
then give a specific example to show  $\Pr(\bar{\x}\neq\hbx|\bar{\x}\in \mathcal{B})\geq 0.9275$ and finally perform some
Matlab simulations to illustrate this example.

\begin{theorem} \label{th:conj}
For any given  $\epsilon>0$, any fixed dimension $n\geq2$, any box $\mathcal{B}$ and any standard deviation $\sigma$ of the noise vector $\v$, there always exists a box-constrained linear model in the form of \eqref{e:lmqr}, where $\hbx$ is uniformly distributed over the box $\mathcal{B}$,  such that
\begin{align}
\label{ine:conj}
\Pr(\bar{\x}\neq\hbx|\bar{\x}\in \mathcal{B})\geq 1-\frac{1}{u_1-\ell_1+1}-\epsilon.
\end{align}
\end{theorem}

{\bf Proof}. Note that
\begin{align}
\label{e:conj}
\Pr(\bar{\x} \neq\hat{\x}|\bar{\x}\in \mathcal{B})
= & \ \frac{\Pr(\bar{\x}\in \mathcal{B})-\Pr(\bar{\x}=\hat{\x}, \bar{\x}\in \mathcal{B})}{\Pr(\bar{\x}\in \mathcal{B})} \nonumber \\
= & \ 1-\frac{\Pr(\bar{\x}=\hat{\x})}{\Pr(\bar{\x}\in \mathcal{B})}.
\end{align}
where the second equality is due to the fact that $\hbx\in \mathcal{B}$.
Thus,  to prove the theorem,  it suffices to show that for any given $\epsilon>0$
there exists a box-constrained linear model such that
\begin{align} \label{ine:pbx}
\frac{\Pr(\bar{\x}=\hat{\x})}{ \Pr(\bar{\x}\in \mathcal{B})} \leq  \frac{1}{u_1-\ell_1+1}+\epsilon.
\end{align}

For any fixed $\sigma$ and $\mathcal{B}$,
to construct the linear model, we need only to construct a matrix $\R\in \Rbb^\mbn$.
Define
$$
\R=\bmx r_{11}&0.5r_{11}\e^T \\ \0  & r_{22}\I_{n-1,n-1}\emx, \ \ 0 < r_{11} \leq r_{22}
$$
where $\e=[1,\ldots,1]^T \in \Rbb^{n-1}$.
We will show how to choose $r_{11}$ such that \eqref{ine:pbx} holds.

Note that $\R$ is already LLL reduced, thus,
$\bar{\x}=\z^\sOB=\x^\sOB$ and $\hbx=\hbz$.
Then, by  \eqref{e:pob} and \eqref{e:varphi},
\beq \label{e:pbx}
\Pr(\bar{\x}=\hat{\x})=\phi_\sigma(r_{11})
\left(\phi_\sigma(r_{22})\right)^{n-1}.
\eeq
Obviously, with event $E_2 \equiv (x_2^\sOB=\hat{x}_2,\ldots, x_n^\sOB=\hat{x}_n)$,
\begin{align}
\Pr(\bar{\x}\in \mathcal{B})
& = \Pr(\x^\sOB\in \mathcal{B}) \geq \Pr\left(x_1^\sOB \in [\ell_1,u_1], E_2 \right)\nonumber \\
&= \Pr (x_1^\sOB \in [\ell_1,u_1]| E_2) \cdot \Pr (E_2) \nonumber \\
& =  \Pr (x_1^\sOB \in [\ell_1,u_1]| E_2) \left(\phi_\sigma(r_{22})\right)^{n-1}
\label{e:pxb}
\end{align}
where  the last equality follows from \eqref{e:pob} and \eqref{e:varphi}.
Therefore, by \eqref{e:pbx} and \eqref{e:pxb}, to show \eqref{ine:pbx} it suffices to show that there exists an $r_{11}>0$ such that
\begin{align} \label{e:pbx1}
\frac{\phi_\sigma(r_{11})}{\Pr(x_1^\sOB \in [\ell_1,u_1]|E_2)}
\leq  \frac{1}{u_1-\ell_1+1}+\epsilon\Big.
\end{align}

In the following we derive an expression for $\Pr(x_1^\sOB \in [\ell_1,u_1]|E_2)$
and then use it to show that \eqref{e:pbx1} holds for some $r_{11}>0$.
From the proof of Theorem \ref{t:pmfpb},
we see that if $x_i^\sOB=\hat{x}_i$ for $i=n,n-1, \ldots, 2$ and $\hx_1$ is fixed, then (cf.\ \eqref{e:dis})
\beq
\label{e:cobdis}
\frac{(c_1^\sOB-\hx_1)r_{11}}{\sqrt{2}\sigma}\sim {\cal N}\left(0, \frac{1}{2}\right).
\eeq
Since  $x_1^\sOB=\lfloor c_1^\sOB \rceil$ and $\hbx$ is uniformly distributed over the box $\mathcal{B}$,
\begin{align*}
&\Pr\big(x_1^\sOB \in [\ell_1,u_1]|E_2 \big) \nonumber \\
 =& \sum_{i=0}^{u_1-\ell_1}\Pr\!\big(\hat{x}_1=\ell_1+i,x_1^\sOB \in [\ell_1,u_1]|E_2 \big) \nonumber \\
 =&\sum_{i=0}^{u_1-\ell_1}\Pr(\hat{x}_1=\ell_1+i)\\
 &\Pr\!\big(x_1^\sOB \in [\ell_1,u_1]|(\hat{x}_1=\ell_1+i, E_2)\big) \nonumber \\
 =& \frac{1}{u_1-\ell_1+1}\times\nonumber \\
&\sum_{i=0}^{u_1-\ell_1}\Pr\!\big(c_1^\sOB \in [\ell_1-1/2, u_1+1/2]| (\hat{x}_1=\ell_1+i, E_2) \big) \nonumber\\
 =& \frac{1}{u_1-\ell_1+1}\sum_{i=0}^{u_1-\ell_1}\Pr\!
 \left(\frac{(c_1^\sOB-\hx_1)r_{11}}{\sqrt{2}\sigma} \right.\nonumber \\
 &\in
 \left.\Big[\frac{-(2i+1)r_{11}}{2\sqrt{2}\sigma},\frac{(2u_1-2\ell_1-2i+1)r_{11}}{2\sqrt{2}\sigma}\Big]\Big| E_2 \right)\nonumber  \\
 =& \frac{1}{2(u_1-\ell_1+1)}\sum_{i=0}^{u_1-\ell_1}
\left[\phi_\sigma((2u_1-2\ell_1-2i+1)r_{11})\right.\nonumber \\
&
 \left.+\phi_\sigma((2i+1)r_{11})\right]
%\label{e:prxob}
\end{align*}
where the second equality follows from \eqref{e:probabc}, and the last equality is due to \eqref{e:varphi} and \eqref{e:cobdis}.

It is easy to verify by L'H\^{o}pital's rule that
\begin{align*}
\lim_{r_{11}\rightarrow0}\frac{\phi_\sigma(r_{11})}{\alpha}=\frac{1}{u_1-\ell_1+1},
\end{align*}
where
\begin{align*}
\alpha=&\frac{1}{2(u_1-\ell_1+1)}\sum_{i=0}^{u_1-\ell_1}
\left[\phi_\sigma((2u_1-2\ell_1-2i+1)r_{11})\right.\nonumber \\
&\quad
 \left.+\phi_\sigma((2i+1)r_{11})\right].
\end{align*}

Therefore, for any $\epsilon>0$,  there exists an $r_{11}>0$ such that \eqref{e:pbx1} holds,
completing the proof.  \ \ $\Box$
\medskip

As $u_1-\ell_1 +1$ is at least 2, Theorem \ref{th:conj} shows that $\Pr(\bar{\x}\neq\hbx|\bar{\x}\in \mathcal{B})$
can be at least $1/2-\epsilon$.
Note that $u_1-\ell_1+1$  can be arbitrarily large and
we can choose $\ell_1$, $u_1$ and $r_{11}$
such that  $\Pr(\bar{\x}\neq\hbx|\bar{\x}\in \mathcal{B})$ can be arbitrarily close  to 1.
In the following, we give a specific example
 to show that $\Pr(\bar{\x}\neq\hbx|\bar{\x}\in \mathcal{B})\geq 0.9275$
 and give some simulation results.

\begin{example}\label{ex:conj}
For any fixed $n$ and $\sigma$, let $\epsilon=0.01$ and $\mathcal{B}=[0, 15]^n$,  and define
\begin{align}
\label{e:exA}
\R=\bmx 0.04 \sigma &0.02 \sigma\e^T \\ \0  & 10 \sigma\I_{n-1,n-1}\emx.
\end{align}
It is easy to verify that this matrix $\R$ satisfies the conditions given in the proof of Theorem \ref{th:conj}.
%In fact, \eqref{ine:r11} holds as the left and right-hand sides of \eqref{ine:r11} are $0.0160$ and $0.0179$, respectively.
Then by \eqref{ine:conj}, we have  $\Pr(\bar{\x}\neq\hbx|\bar{\x}\in \mathcal{B})\geq 0.9275$.

We use \textsc{MATLAB} to do some simulations to illustrate  the probability.
In the simulations, for any fixed $n$ and $\sigma$, we generated an $n\times n$ matrix $\R$ by using \eqref{e:exA}.
After fixing $\R$, we gave 10000 runs to generate 10000 pairs of $\hbx$ and $\tilde{\v}$ according to their distributions,
producing 10000 $\tby$'s according to  \eqref{e:modelqr}.
For each $\tby$,  we found the Babai point $\x^\sOB$  by using \eqref{e:BabaiO}.
For each pair of $\R$ and $\sigma$, we computed the theoretical  probability $\Pr(\bar{\x}\neq\hat{\x})$ denoted by $P_{err}$
by using \eqref{e:pbx} (notice that $P_{err}=1-P^\sOB$ since $\bbx=\x^\sOB$ here)
and the corresponding experimental probability $P_{ex}$ (i.e., the ratio of the number of runs in which $\bar{\x}\neq\hbx$ to 10000).
We also computed the experimental probability $P_b$ of $\bar\x \in\mathcal{B}$ (i.e., the ratio of the number of runs in which
$\bar\x \in\mathcal{B}$ to 10000)
and the experimental probability $P_e$ corresponding to $\Pr(\bar{\x}\neq\hbx|\bar{\x}\in \mathcal{B})$,
i.e., $P_e =1 -(1-P_{ex})/P_{b}$ (cf.\ \eqref{e:conj}).

Tables \ref{tb:tbconj1} and \ref{tb:tbconj2} respectively display those probabilities versus $n=5:5:40$ with $\sigma=0.1$ and
versus $\sigma=0.1:0.1:0.8$ with $n=20$.
From these two tables, we can see that the values of $P_e$ are larger than 0.9275 except the case that $n=40$ in Table \ref{tb:tbconj1},
in which $P_e$ is smaller than 0.9275, but it is close to the latter.
Thus the test results are consistent with the theoretical result.
We also observe that $P_{th}$ is very small and  $P_{ex}$ is a good approximation to $P_{th}$.
In Tables \ref{tb:tbconj1} the values of $P_{err}$  are actually different,
but very close  because $\phi_\sigma(r_{22})$ is very close to 1 (c.f.\ \eqref{e:pbx})
and in Tables \ref{tb:tbconj2}  the values of $P_{err}$ are exactly equal because in \eqref{e:pbx}
$\phi_\sigma(r_{11})$ and
 $\phi_\sigma(r_{22})$ are independent of $\sigma$.
This experiment confirms that even if $\bbx$ is valid, there may be a large chance that it is in error.

%We also observe that $P_{ex}$ is always less than $P_b$, indeed this is true because $P_{th}\leq P_b$ in theory.

\begin{table}[h!]
\caption{Probabilities versus $n=5:5:40$ with $\sigma=0.1$}
\centering
\begin{tabular}{|c||c|c|c|c|c|}
\hline
$\n$ & $P_{err}$ & $P_{ex}$ & $P_b$ & $P_e$ \\ \hline
5 & 0.9840  & 0.9840 & 0.2484 & 0.9356   \\ \hline
10 & 0.9840 & 0.9850 & 0.2485 & 0.9396   \\ \hline
15 & 0.9840 & 0.9836 & 0.2469 & 0.9336   \\ \hline
20 & 0.9840 & 0.9837 & 0.2499 & 0.9348   \\ \hline
25 & 0.9840 & 0.9827 & 0.2564 & 0.9325   \\ \hline
30 & 0.9840 & 0.9828 & 0.2500 & 0.9312   \\ \hline
35 & 0.9840 & 0.9828 & 0.2473 & 0.9304   \\ \hline
40 & 0.9840 & 0.9818 & 0.2434 & 0.9252   \\ \hline
\end{tabular}
\label{tb:tbconj1}
\end{table}

\begin{table}[h!]
\caption{Probabilities versus $\sigma=0.1:0.1:0.8$ with $n=20$}
\centering
\begin{tabular}{|c||c|c|c|c|c|}
\hline
$\sigma$ & $P_{err}$ & $P_{ex}$ & $P_b$ & $P_e$  \\ \hline
0.1 & 0.9840 & 0.9841 & 0.2503 & 0.9365   \\ \hline
0.2 & 0.9840 & 0.9832 & 0.2522 & 0.9334   \\ \hline
0.3 & 0.9840 & 0.9844 & 0.2434 & 0.9359   \\ \hline
0.4 & 0.9840 & 0.9844 & 0.2399 & 0.9350   \\ \hline
0.5 & 0.9840 & 0.9825 & 0.2435 & 0.9281   \\ \hline
0.6 & 0.9840 & 0.9827 & 0.2475 & 0.9301   \\ \hline
0.7 & 0.9840 & 0.9843 & 0.2541 & 0.9382   \\ \hline
0.8 & 0.9840 & 0.9838 & 0.2517 & 0.9356   \\ \hline
\end{tabular}
\label{tb:tbconj2}
\end{table}
\end{example}

\subsection{The conjecture holds under some conditions}

In this subsection, we will show that the conjecture holds under some conditions.

Recall that $\bar{\x}=\Z \z^\sOB$ and $\hbx=\Z \hbz$, thus
$\Pr(\bar{\x}=\hbx)=\Pr(\z^\sOB=\hbz)$.
Then, by \eqref{e:conj}, we have
\begin{align}
\label{ine:conjbd}
\Pr(\bar{\x}\neq\hbx|\bar{\x}\in \mathcal{B})\leq 1-\Pr(\z^\sOB=\hat{\z}).
\end{align}
So, if $\Pr(\z^\sOB=\hat{\z})\approx 1$, then the conjecture holds.
%For example, by the simulation (Table IV) in \cite{ChaWX12}, if $n=20$, the
%entries of $\A$ independently follow the normal distribution and $\sigma\leq0.35$, then $\Pr(\z^\sOB=\hat{\z})\geq 0.97589$.
%(\textcolor{red}{The result holds for a fixed $\A$? Note that we assume $\A$ is fixed in our model})
From Corollary \ref{c:Babai1} we see that when $\sigma$ is small enough,
we have $\Pr(\z^\sOB=\hat{\z})\approx 1$.
But the upper bound given in \eqref{ine:conjbd} is not sharp because it was derived from \eqref{e:conj} by using
the inequality  $\Pr(\bar{\x}\in \mathcal{B})\leq 1$.
We will give a sharper upper bound on $\Pr(\bar{\x}\neq\hbx|\bar{\x}\in \mathcal{B})$
based on a sharper upper bound on  $\Pr(\bar{\x}\in \mathcal{B})$.

Since $\bar{\x}=\Z\z^\sOB$, $\bar{\x}\in \mathcal{B}$ if and only if $\z^\sOB\in \mathcal{E}
\equiv \{\Z^{-1}\s \,|\, \forall  \s \in \mathcal{B}\}$.
Thus $\Pr(\bar{\x}\in \mathcal{B})=\Pr(\z^\sOB\in \mathcal{E})$.
But   the set $\mathcal{E}$ is a parallelotope and it is difficult to analyze $\Pr(\z^\sOB\in \mathcal{E})$.
Thus in the following we will give a box $\mathcal{F}$ which contains $\mathcal{E}$, then we analyze $\Pr(\z^\sOB\in \mathcal{F})$.
Let $\U=(u_{ij})=\Z^{-1}$ and
define   for $i,j=1,2,\ldots, n$,
\beqnn
\begin{split}
& \mu_{ij}=
\begin{cases}
\ell_j, & \mbox{ if }\   u_{ij} \geq 0\\
u_j, & \mbox{ if }\    u_{ij}<0
\end{cases}
\end{split}, \  \
\begin{split}
& \nu_{ij}=
\begin{cases}
u_j, & \mbox{ if }\   u_{ij} \geq 0\\
\ell_j, & \mbox{ if }\    u_{ij}<0
\end{cases}
\end{split}.
\eeqnn
Then define $\bar{\l}\in \Zn$ and $\bar{\u}\in \Zn$ as follows:
\begin{align}
\label{e:boxLLL}
\bar{\ell}_i=\sum_{j=1}^nu_{ij}\mu_{ij}, \ \ \bar{u}_i=\sum_{j=1}^nu_{ij}\nu_{ij}, \ \ i=1,2,\ldots,n.
\end{align}
It is easy to observe that
\beq \label{e:boxF}
\mathcal{E}\subseteq \mathcal{F} \equiv \{\z\in \mathbb{Z}^n:\bar{\l}\leq \z\leq\bar{\u}\}.
\eeq
Actually it is easy to observe that $\mathcal{F}$ is the smallest box including $\mathcal{E}$.

With the above preparation, we now give the following result.

\begin{theorem} \label{t:valid}
Suppose that the assumptions in Theorem \ref{t:pmfpb} hold
and the linear model \eqref{e:modelqr} is transformed to the linear model \eqref{e:modelC} through the LLL reduction \eqref{e:LLLC}.
Then the estimator $\bbx$ defined as $\bbx=\Z\z^\sOB$ satisfies
\begin{align}\label{ine:conjbd2}
&\Pr(\bar{\x}\neq\hbx|\bar{\x}\in \mathcal{B})\nonumber\\
\leq& 1-\prod_{i=1}^n
\frac{\mbox{erf}\left(\br_{ii}/(2\sqrt{2}\sigma)\right)}
{\mbox{erf}\left((\bar{u}_i-\bar{\ell}_i+1)\bar{r}_{ii}/(2\sqrt{2}\sigma)\right)}
\end{align}
where $\bbl$ and $\bbu$ are defined in \eqref{e:boxLLL}.
\end{theorem}

{\bf Proof}. Since $\Pr(\bar{\x}\in \mathcal{B})=\Pr(\z^\sOB\in \mathcal{E})$, it follows from \eqref{e:boxF} that
\begin{align}
\label{e:bdF}
\Pr(\bar{\x}\in \mathcal{B})\leq\Pr(\z^\sOB\in \mathcal{F}).
\end{align}
In the following, we will show
\beq
\Pr(\z^\sOB\in \mathcal{F})\leq\prod_{i=1}^n\phi_\sigma((\bar{u}_i-\bar{\ell}_i+1)\bar{r}_{ii})
\label{e:pzobub}
\eeq
where $\phi_\sigma(\cdot)$ is defined in \eqref{e:varphi}.
Then combining \eqref{e:bdF},  \eqref{e:pzobub} and
the fact that $\Pr(\bbx=\hbx)=\Pr(\z^\sOB=\hbz)=\prod_{i=1}^n\phi_\sigma(\br_{ii})$
(see \eqref{e:pob} and \eqref{e:varphi}),
we can conclude that \eqref{ine:conjbd2} holds from \eqref{e:conj}.

To show \eqref{e:pzobub},  instead of analyzing the probability
of $\z^\sOB$ on its left-hand side, we will analyze an equivalent probability of $\bar{\v}$
as we know its distribution.

In our proof, we need to use the basic result: given $ v \sim \mathcal{N}(0, \sigma^2)$ and  $\eta > 0$,
for any $\zeta \in \mathbb{R}$,
\beq  \label{e:gauss}
\Pr(v \in[\zeta , \zeta +\eta])\leq \Pr(v \in[-\eta/2,\eta/2])=\phi_\sigma(\eta).
\eeq
By \eqref{e:varphi}, the equality in \eqref{e:gauss} obviously holds.
In the following, we show the inequality holds, i.e., equivalently show
\[
\int_{\zeta}^{\zeta+\eta}\exp\big(-\frac{t^2}{2\sigma^2}\big)dt
\leq\int_{-\eta/2}^{\eta/2}\exp\big(-\frac{t^2}{2\sigma^2}\big)dt.
\]
For any fixed $\eta>0$, the left-hand side of the above inequality is a function of $\zeta$.
Furthermore, it can be easily verified that the derivative of this function equal to zero if  $\zeta=-\eta/2$,
and it is positive when $\zeta<-\eta/2$ and negative when $\zeta>-\eta/2$.
Thus, this function achieves the maximal value when $\zeta=-\eta/2$, so the inequality holds.

From \eqref{e:modelC} and \eqref{e:BabaiC},
we observe that $\z^\sOB$ is a function of $\bar{\v}$.
To emphasize this, we write it as $\z^\sOB(\bar{\v})$.
When $\bar{\v}$ changes, $\z^\sOB(\bar{\v})$ may change too.
In the following analysis, we assume that $\hbz$  is fixed
and $\bar{\v}$ satisfies the model \eqref{e:modelC}.

For later uses, for $=n,n-1,\ldots,1$, define
\begin{align}
\mathcal{G}_i=\{\w_{i:n}&|\,  \bar{\y}_{i:n} = \bar{\R}_{i:n, i:n}\hbz_{i:n} + \w_{i:n}, \,
\w_{i:n} \in \Rbb^{n-i+1},\, \nonumber\\
&z^\sOB_k(\w_{k:n})\in [\bar{\ell}_k, \bar{u}_k], \,k=i, i+1, \ldots,  n\}.
\label{e:Fi}
\end{align}
From \eqref{e:Fi}, it is easy to verify that
%for any $\bar{\v}$ satisfies \eqref{e:modelC}, $\z^\sOB(\bar{\v})$ computed by \eqref{e:BabaiC} satisfies
$\z^\sOB(\bar{\v})\in \mathcal{F}$ if and only if $\bar{\v}\in \mathcal{G}_1$.
Therefore,   \eqref{e:pzobub} is equivalent to
\begin{align}
\label{e:F1}
\Pr(\bar{\v}_{1:n}\in \mathcal{G}_1)\leq\prod_{i=1}^n\phi_\sigma((\bar{u}_i-\bar{\ell}_i+1)\bar{r}_{ii}).
\end{align}

We prove \eqref{e:F1} by induction. First, we prove the base case:
\begin{align*}
\Pr(\bar{v}_{n}\in \mathcal{G}_n)\leq\phi_\sigma((\bar{u}_n-\bar{\ell}_n+1)\bar{r}_{nn}).
\end{align*}
By  \eqref{e:BabaiC} and \eqref{e:modelC}, we have
\begin{align*}
c_{n}^\sOB= \frac{\bar{y}_{n}}{\bar{r}_{nn}}
= \frac{\bar{r}_{nn}\hat{z}_{n}+\bar{v}_{n}}{\bar{r}_{nn}}
=\hz_{n}+\frac{\bar{v}_{n}}{\bar{r}_{nn}}.
\end{align*}
Since $z^\sOB_n(\bar{v}_{n})=\lfloor c_{n}^\sOB\rceil$, by \eqref{e:Fi},
\begin{align*}
&\Pr(\bar{v}_{n}\in \mathcal{G}_n)\\
=&\Pr\Big( \hz_{n}+ \frac{\bar{v}_{n}}{\bar{r}_{nn}}\in [\bar{\ell}_n-1/2,\bar{u}_n+1/2]\Big) \nonumber  \\
 =&\Pr(\bar{v}_{n}\in[(\bar{\ell}_n-\hat{z}_{n}-1/2)\bar{r}_{nn}, (\bar{u}_n-\hat{z}_{n}+1/2)\bar{r}_{nn}]) \nonumber  \\
\leq&\phi_\sigma((\bar{u}_n-\bar{\ell}_n+1)\bar{r}_{nn}),
\end{align*}
where the inequality follows from \eqref{e:gauss}.

Suppose for some $i >1$, we have
\beq
\label{e:PFi}
\Pr(\bar{\v}_{i:n}\in \mathcal{G}_i)\leq\prod_{k=i}^n\phi_\sigma((\bar{u}_k-\bar{\ell}_k+1)\bar{r}_{kk}).
\eeq
Now we want to prove
\beq
\label{e:Fi-1}
\Pr(\bar{\v}_{i-1:n}\in \mathcal{G}_{i-1})\leq\prod_{k=i-1}^n\phi_\sigma((\bar{u}_k-\bar{\ell}_k+1)\bar{r}_{kk}).
\eeq

We  partition the set $\mathcal{G}_i$ into a sequence of disjoint subsets.
To do that, for $i=n,n-1,\ldots, 1$, we first define  the discrete set
\begin{align*}
\mathcal{H}_i=\Big\{ \sum_{j=i}^n\frac{r_{i-1,j}}{r_{i-1,i-1}}(\hat{z}_j-z^\sOB_j(\w_{j:n}))\big| \w_{i:n}\in \mathcal{G}_i\Big\}.
\end{align*}
Then, for any $t\in \mathcal{H}_i$, we define
\begin{align*}
\mathcal{G}_{i,t}
=\Big\{\w_{i:n}&|\w_{i:n}\in \mathcal{G}_i \mbox{ such that }\Big.\\
&\Big.\sum_{j=i}^n\frac{r_{i-1,j}}{r_{i-1,i-1}}(\hat{z}_j-z^\sOB_j(\w_{j:n}))=t\Big\}.
\end{align*}
It is easy to verify that $\cup_{t\in \mathcal{H}_i}\mathcal{G}_{i,t}=\mathcal{G}_{i}$ and
$\mathcal{G}_{i,t_1}\cap \mathcal{G}_{i,t_2}=\varnothing$ for $t_1, t_2\in \mathcal{H}_{i}$ and $t_1\neq t_2$.
Therefore,
\beq
\Pr(\bar{\v}_{i:n}\in \mathcal{G}_i)
= \sum_{t\in\mathcal{H}_i} \Pr(\bar{\v}_{i:n}\in \mathcal{G}_{i,t})
\label{e:probi}
\eeq
and
\begin{align}
&\Pr(\bar{\v}_{i-1:n}\in \mathcal{G}_{i-1})\nonumber\\
=& \Pr(\bar{\v}_{i:n}\in \mathcal{G}_{i},\,  z^\sOB_{i-1}(\bar{\v}_{i-1:n})\in [\bar{\ell}_{i-1}, \bar{u}_{i-1}])
\nonumber \\
=& \sum_{t\in\mathcal{H}_i}\Pr(\bar{\v}_{i:n}\in \mathcal{G}_{i,t},\,  z^\sOB_{i-1}(\bar{\v}_{i-1:n})\in [\bar{\ell}_{i-1}, \bar{u}_{i-1}])
\nonumber \\
=&  \sum_{t\in\mathcal{H}_i}\Pr(\bar{\v}_{i:n}\in \mathcal{G}_{i,t})\nonumber\\
&\times\Pr\left(z^\sOB_{i-1}(\bar{\v}_{i-1:n})\in [\bar{\ell}_{i-1}, \bar{u}_{i-1}]|\bar{\v}_{i:n}\in \mathcal{G}_{i,t}\right).
\label{e:Fomega}
\end{align}

Now we  derive a bound on the second probability of each term on the right-hand side of \eqref{e:Fomega}.
By  \eqref{e:BabaiC} and \eqref{e:modelC}, we have
%\begin{align*}
%c_{i-1}^\sOB
%=&\frac{\bar{r}_{i-1,i-1}\hat{z}_{i-1}+\sum_{j=i}^n\bar{r}_{i-1,j}(\hat{z}_j-z^\sOB_j(\bar{\v}_{j:n}))
%+\bar{v}_{i-1}}{\bar{r}_{i-1,i-1}}\\
%=&\hz_{i-1}+ t'+\frac{\bar{v}_{i-1}}{\bar{r}_{i-1,i-1}}
%\end{align*}
\begin{align*}
c_{i-1}^\sOB
=&\frac{\bar{r}_{i-1,i-1}\hat{z}_{i-1}}{\bar{r}_{i-1,i-1}}
+\frac{\sum_{j=i}^n\bar{r}_{i-1,j}(\hat{z}_j-z^\sOB_j(\bar{\v}_{j:n}))}{\bar{r}_{i-1,i-1}}\\
&+\frac{\bar{v}_{i-1}}{\bar{r}_{i-1,i-1}}\\
=&\hz_{i-1}+ t'+\frac{\bar{v}_{i-1}}{\bar{r}_{i-1,i-1}}
\end{align*}
where
$$
t'=\sum_{j=i}^n\frac{\bar{r}_{i-1,j}}{\bar{r}_{i-1,i-1}}(\hz_j-z^\sOB_j(\bar{\v}_{j:n})).
$$
If $\bar{\v}_{i:n}\in \mathcal{G}_{i,t}$, $t'\in \mathcal{H}_{i}$.
Since   $z^\sOB_{i-1}(\bar{\v}_{i-1:n})=\lfloor c_{i-1}^\sOB\rceil$,
\begin{align}
&\Pr\left(z^\sOB_{i-1}(\bar{\v}_{i-1:n})\in [\bar{\ell}_{i-1}, \bar{u}_{i-1}]|\bar{\v}_{i:n}\in \mathcal{G}_{i,t}\right) \nonumber  \\
= \ & \Pr\Big(\hz_{i-1}+t' +\frac{\bar{v}_{i-1}}{\bar{r}_{i-1,i-1}}\in [\bar{\ell}_{i-1}-1/2,\bar{u}_{i-1}+1/2]\Big)  \nonumber \\
=  \ & \Pr\left(\bar{v}_{i-1}\in \big[(\bar{\ell}_{i-1}-\hz_{i-1}-t'-1/2)\bar{r}_{i-1,i-1}, \right.\,\nonumber \\
   &\left. (\bar{u}_{i-1}-\hz_{i-1} -t'+1/2)\bar{r}_{i-1,i-1}\big]\right)
\nonumber \\
\leq \ & \phi_\sigma((\bar{u}_{i-1}-\bar{\ell}_{i-1}+1)\bar{r}_{i-1,i-1})
\label{e:probim1b}
\end{align}
where the inequality follows from \eqref{e:gauss}.
Thus, from \eqref{e:Fomega} it follows that
\begin{align*}
&\Pr(\bar{\v}_{i-1:n}\in \mathcal{G}_{i-1})\\
\leq& \sum_{t\in\mathcal{H}_i}\Pr(\bar{\v}_{i:n}\in \mathcal{G}_{i,t}) \phi_\sigma((\bar{u}_{i-1}-\bar{\ell}_{i-1}+1)\bar{r}_{i-1,i-1}) \\
= &\Pr(\bar{\v}_{i:n}\in \mathcal{G}_{i}) \phi_\sigma((\bar{u}_{i-1}-\bar{\ell}_{i-1}+1)\bar{r}_{i-1,i-1})
\end{align*}
where the equality is due to \eqref{e:probi}.
Then the inequality \eqref{e:Fi-1} follows by using the induction hypothesis  \eqref{e:PFi}.
Therefore,  the inequality \eqref{e:F1}, or the equivalent inequality \eqref{e:pzobub}, holds for any fixed $\hat{\z}$.

Since \eqref{e:pzobub} holds for any fixed $\hat{\z}$,
it is easy  to argue that it holds no matter what distribution of $\hat{\z}$ is over the box $\mathcal{F}$,
so the theorem is proved.\ \ $\Box$

\medskip

By Theorem \ref{t:valid},
if $\prod_{i=1}^n\frac{\phi_\sigma(\bar{r}_{ii})}{\phi_\sigma((\bar{u}_i-\bar{\ell}_i+1)\bar{r}_{ii})} \approx 1$,
then $\Pr(\bar{\x}\neq\hbx|\bar{\x}\in \mathcal{B})\approx0$, i.e., the conjecture holds.
Again, the condition will be satisfied when the noise standard deviation $\sigma$ is sufficiently small.
Simulations in \cite{MaHFL09} showed that for practical MIMO systems often $\Pr(\bar{\x}\neq\hbx|\bar{\x}\in \mathcal{B})\approx0$.

Here we  make a comment on the upper bound in \eqref{ine:conjbd2}.
The derivation of  \eqref{ine:conjbd2} was based on the two inequalities  \eqref{e:bdF} and \eqref{e:pzobub}.
The inequality \eqref{e:bdF} was established based on  the fact that $\mathcal{E} \subseteq \mathcal{F}$ in \eqref{e:boxF}.
If the absolute values of the entries of the unimodular matrix $\Z^{-1}$ are big, then it is likely that $\mathcal{F}$ is much bigger than $\mathcal{E}$
although  $\mathcal{F}$ is the smallest box containing $\mathcal{E}$, making
the inequality \eqref{e:bdF}  loose.
Otherwise it will be tight; in particular, when $\Z=\I$, then $\mathcal{E}=\mathcal{F}$
and  the inequality \eqref{e:bdF} becomes an equality.
In establishing the inequality \eqref{e:pzobub} we used the inequality \eqref{e:gauss} (see \eqref{e:probim1b}),
which is simple but may not be tight if $\zeta$ is not close to $-\eta/2$.
Thus the inequality \eqref{e:pzobub} may not be tight.
Overall, the upper bound in \eqref{ine:conjbd2} may not be tight sometimes,
but it is always tighter than the upper bound given by  \eqref{ine:conjbd}.
The following  example shows  that the former can be significantly tighter than the latter
and can be a sharp bound.

\begin{example}\label{ex:conjbd}
We use exactly the same data generated in Example \ref{ex:conj} to compute  the upper bounds in \eqref{ine:conjbd}  and \eqref{ine:conjbd2},
which are denoted by $\mu_{eb1}$ and $\mu_{eb2}$, respectively.
The results for $n=5:5:40$ with $\sigma=0.1$ are given in Table \ref{tb:tbconjbd1}.
To see how tight  they are,  the values of $P_{e}$ given in Table \ref{tb:tbconj1} are displayed here again.
Recall  $P_{e}$ is  the experimental probability corresponding to the theoretical probability
$\Pr(\bbx \neq \hbx|\bbx\in \mathcal{B})$  in    \eqref{ine:conjbd} and \eqref{ine:conjbd2}.

From Table \ref{tb:tbconjbd1},  we can see the upper bound $\mu_{eb2}$ is obviously tighter than  the upper bound  $\mu_{eb1}$
and $\mu_{eb2}$ is close to $P_{e}$.
When $n=10$, $P_{e}>\mu_{eb2}$, this is because there are some deviations between the experimental values and the theoretical values.
The values of $\mu_{eb1}$ are actually not exactly the same for different $n$, but they are very close.
This is also true for $\mu_{eb2}$.

\begin{table}[h!]
\caption{$P_e$ and bounds versus $n=5:5:40$ with $\sigma=0.1$}
\centering
\begin{tabular}{|c||c|c|c|c|c|}
\hline
$\n$ &  $P_{e}$ & $\mu_{eb1}$ & $\mu_{eb2}$ \\ \hline
5 &  0.9356 & 0.9840 & 0.9364   \\ \hline
10  & 0.9396 & 0.9840 & 0.9364   \\ \hline
15  & 0.9336 & 0.9840 & 0.9364   \\ \hline
20  & 0.9348 & 0.9840 & 0.9364   \\ \hline
25  & 0.9325 & 0.9840 & 0.9364   \\ \hline
30  & 0.9312 & 0.9840 & 0.9364   \\ \hline
35  & 0.9304 & 0.9840 & 0.9364   \\ \hline
40  & 0.9252 & 0.9840 & 0.9364   \\ \hline
\end{tabular}
\label{tb:tbconjbd1}
\end{table}
\end{example}

\section{Summary and future work} \label{s:sum}

We have presented formulas for the success probability $P^\sBB$ of the box-constrained Babai estimator  and
the success probability $P^\sOB$ of the ordinary Babai estimator  for the linear model where the true integer
parameter vector $\hbx$  is  uniformly distributed over the  constraint box and the noise vector follows a normal distribution.
The properties of $P^\sBB$ and $P^\sOB$ and the relationship between them were given.

The effects of the column permutations on $P^\sBB$ by the LLL-P, SQRD and V-BLAST column permutation strategies have been investigated.
When the noise is relatively small, we showed that  LLL-P  always increases $P^\sBB$
and argued why both SQRD and V-BLAST  usually increase $P^\sBB$;
and when the noise is relatively large, LLL-P always decreases $P^\sBB$
and argued why both SQRD and V-BLAST   usually decrease $P^\sBB$.
The latter contradicts with what we commonly believed.
And it suggests that we should check the conditions given in the paper before we apply these  strategies.
We also provided a column permutation invariant bound on $P^\sBB$.
This bound helped us to understand the effects of these column permutation strategies on $P^\sBB$.
Our theoretical findings were supported by numerical test results.

We have given an example to show that the conjecture proposed in [24] does not always hold and imposed a condition under
which  the conjecture holds.

LLL-P has better theory than V-BLAST and SQRD in terms of their effects on $P^\sBB$.
But our numerical experiments indicated often V-BLAST is more effective than LLL-P and SQRD.
Developing a more effective column permutation strategy with solid theory will be investigated in the future.
These three permutation column permutation strategies use only the information of $\A$.
The effects of the  column permutation strategies which use all available information
of the model such as those proposed in  \cite{SuW05},  \cite{ChaH08} and \cite{BreC11} need to be investigated.

Recently the success probability of the BILS estimator has been given in \cite{PapCK13}.
We intend to study the relationship between it and $P^\sBB$.

\section*{Acknowledgment}
We are  grateful to Prof. Frank R. Kschischang and the referees for their valuable and thoughtful suggestions.
Part of this work was undergone while the first author was studying as a Ph.D student at McGill University,
and working as a postdoctoral fellow at Laboratoire de l'Informatique du Parall\'elisme, (CNRS, ENS de Lyon, Inria, UCBL),  Lyon 69007, France,
whose hospitality is gratefully acknowledged.

\bibliographystyle{IEEEtran}
\bibliography{ref}

\end{document}